\documentclass[11pt]{article}
\usepackage{geometry}    
\usepackage{graphicx}
\usepackage{amssymb}
\usepackage{epstopdf}
\usepackage{authblk}
\def\WArP{\sf{WArP}\rm~}

\title{Demonstration and Comparison of Operation of Photomultiplier Tubes at Liquid Argon Temperature}

\author[a]{R.~Acciarri}
\author[b]{M.~Antonello}
\author[c]{F.~Boffelli}
\author[c]{M.~Cambiaghi}
\author[b,*]{N.~Canci}
\author[a]{F.~Cavanna}
\author[d]{A.G.~Cocco}
\author[e,\S]{N.~Deniskina}
\author[a,b,\P]{F.~Di Pompeo}
\author[e]{G.~Fiorillo}
\author[f]{C.~Galbiati}
\author[a,b,$\ddagger$]{L.~Grandi}
\author[g]{P.~Kryczynski}
\author[h]{G.~Meng}
\author[i]{C.~Montanari}
\author[b]{O.~Palamara}
\author[b]{L.~Pandola}
\author[e]{F.~Perfetto}
\author[a]{G.B.~Piano Mortari}
\author[h]{F.~Pietropaolo}
\author[i]{G.L. Raselli}
\author[i]{M.~Rossella}
\author[b]{C.~Rubbia}
\author[b,*]{E.~Segreto}
\author[g,a,$\dagger$,*]{A.M.~Szelc}
\author[j]{A.~Triossi}
\author[h]{S.~Ventura}
\author[b]{C.~Vignoli}
\author[c]{A.~Zani}

\affil[a]{Universit\`a dell'Aquila e INFN, L'Aquila, Italy}
\affil[b]{INFN - Laboratori Nazionali del Gran Sasso, Assergi, Italy}
\affil[c]{Universit\`a di Pavia e INFN, Pavia, Italy}
\affil[d]{INFN - Sezione di Napoli, Napoli, Italy}
\affil[e]{Universit\`a di Napoli e INFN, Napoli, Italy}
\affil[f]{Princeton University - Princeton, New Jersey, USA}
\affil[g]{IFJ PAN, Krakow, Poland}
\affil[h]{INFN - Sezione di Padova, Padova, Italy}
\affil[i]{INFN - Sezione di Pavia, Pavia, Italy}
\affil[j]{INFN - Laboratori Nazionali di Legnaro, Legnaro, Italy}

\date{}                                           

\begin{document}
\maketitle
\let\oldthefootnote\thefootnote
\renewcommand{\thefootnote}{\fnsymbol{footnote}}
\footnotetext[1]{Corresponding authors: ettore.segreto@lngs.infn.it, andrzej.szelc@yale.edu, nicola.canci@lngs.infn.it}
\footnotetext[2]{Currently at Yale University - New Haven, Connecticut, USA.}
\footnotetext[3]{Currently at Princeton University - Princeton, New Jersey, USA.}
\footnotetext[4]{Currently at IASS - Potsdam, Germany.}
\footnotetext[5]{Currently at ITAB - Chieti, Italy.}
\let\thefootnote\oldthefootnote
\begin{abstract}
Liquified noble gases are widely used as a target in direct Dark Matter searches. Signals from scintillation in the liquid, following  energy deposition from the recoil nuclei scattered by Dark Matter particles (e.g. WIMPs), should be recorded down to very low energies by photosensors suitably designed to operate at cryogenic temperatures. Liquid Argon based detectors for Dark Matter searches currently implement photo multiplier tubes for signal read-out. In the last few years PMTs with photocathodes operating down to liquid Argon temperatures (87 K) have been specially developed with increasing Quantum Efficiency characteristics. The most recent of these,  {\sf Hamamatsu Photonics} Mod. R11065 with peak QE up to about 35\%, has been extensively tested within the R\&D program of the \WArP Collaboration. During these testes the Hamamatsu PMTs showed superb performance and allowed obtaining a light yield around 7 {\it phel}/keV$_{ee}$ in a Liquid Argon detector with a photocathodic coverage in the 12\% range, sufficient for detection of events down to few keV$_{ee}$ of energy deposition. This shows that this new type of PMT is suited for experimental applications, in particular for new direct Dark Matter searches with LAr-based experiments.


\end{abstract}


\section{Introduction}
 A new generation, 
high Quantum Efficiency, 3" photomultiplier tube (PMT) for cryogenic applications at liquid Argon temperature (LAr, T=87 K) has been recently developed by  {\sf Hamamatsu Photonics} ({\it Mod. R11065}). This is of interest to experiments adopting liquified Argon as target, in particular for direct Dark Matter searches, and read out the scintillation light signals from interactions in the medium. Using these new PMTs could lead to improvements in detector sensitivity down to low recoil energy thresholds due to their enhanced quantum efficiency. \\
Within the on-going R\&D activity of the \WArP Collaboration a first set of R11065 PMTs has been subject to a series of tests aiming at their characterization in reference working conditions i.e. immersed in liquid argon and optically coupled to LAr cells of various sizes. Scintillation light signals  from interactions in the cell were detected by the PMTs and read out by fast waveform digitizers.\\
A comparison of the R11065 Hamamatsu PMT with a former generation of cryogenic PMT, produced by {\sf Electron Tubes Limited} - Mod. ETL D750 (currently used in the \WArP-100 detector) has been performed by simultaneously operating the two PMTs viewing a common LAr volume. \\
 In these tests the Hamamatsu PMT has shown superb performance demonstrating that it is suited for experimental applications, in particular for new direct Dark Matter searches using LAr-based detectors.

\section{The Hamamatsu PMT}
\begin{figure}[h]
\begin{center}
\includegraphics*[width=11cm]{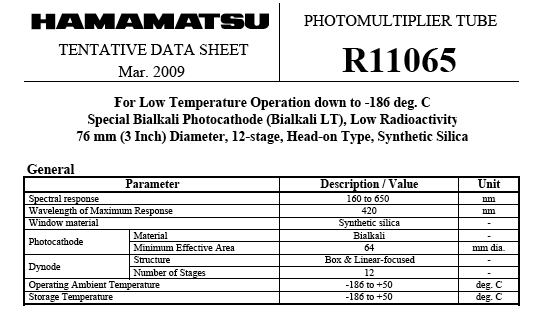}
 \caption{\textsf{\textit{Main characteristics of the Hamamatsu PMT.}}}
\label{fig:hmmts+PMT}
\end{center}
\end{figure}

\begin{figure}[htbp]
\vspace*{-0.5cm}
\begin{center}
\includegraphics*[width=8cm,angle=90]{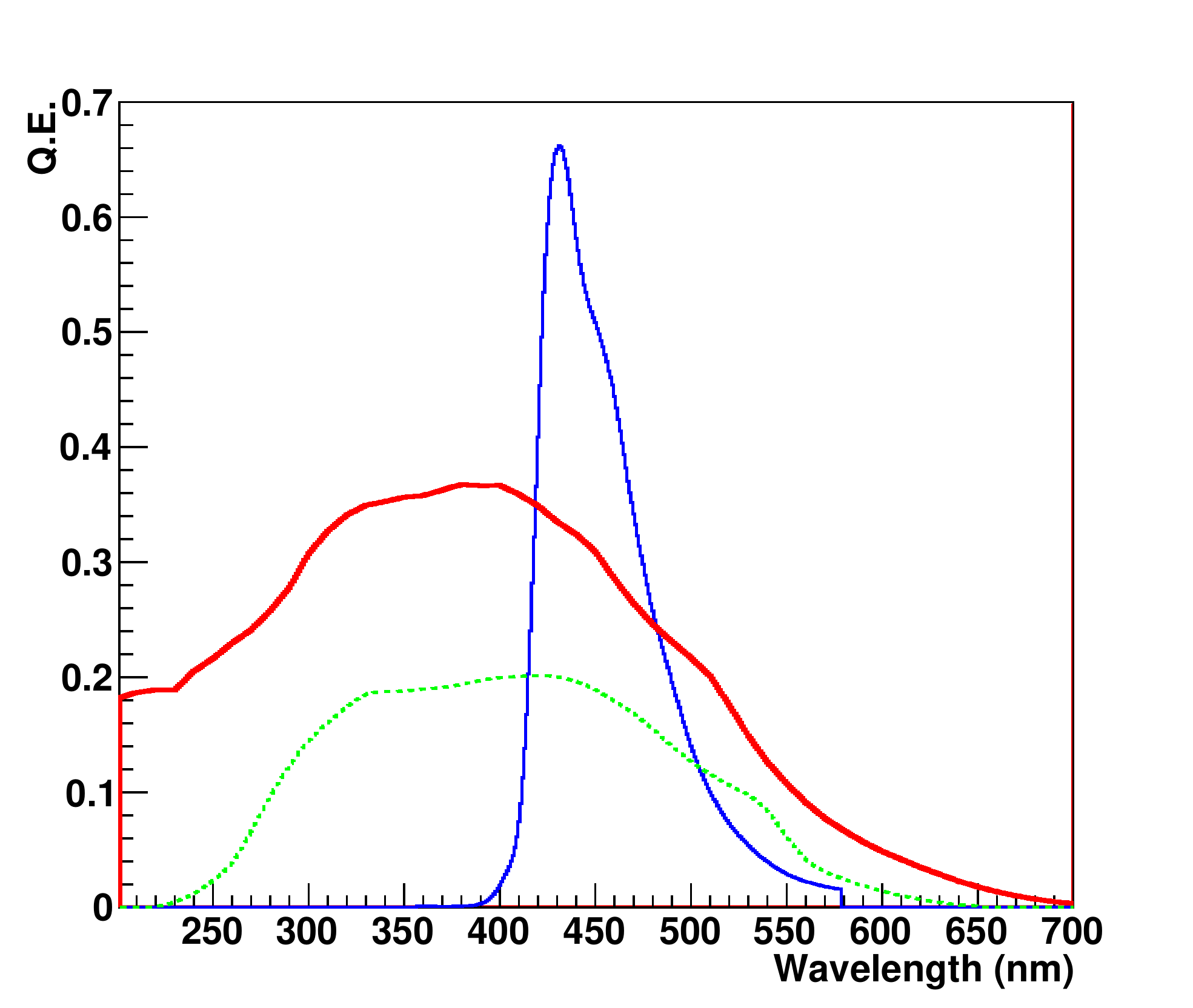}
 \caption{\textsf{\textit{The spectral response of the TPB wavelength-shifter [arb. units, blue line]. Quantum Efficiency of  the R11065 Hamamatsu PMT [red full line], $\langle QE\rangle$ = 29.5\%, averaged over theTPB spectrum.  The Quantum Efficiency of a reference cryogenic PMT (ETL D750) is also reported for comparison [green dotted line], $\langle QE\rangle$ = 17.5\%. }}}
\label{fig:QE}
\end{center}
\end{figure}

The  {\sf Hamamatsu} R11065  \cite{R11065} is a Box\&Linear-focused 12-stages PMT, with Synthetic Silica 3" window (opaque to wavelenghts below 160 nm) and special Bialkali photo-cathode developed to efficiently operate down to LAr temperature in the spectral range from UV to VIS.  
Its prime features include fast time response, good time resolution and pulse linearity.
However, the most noteworthy parameter of this model is its excellent Quantum Efficiency (QE). Hamamatsu declares it at around 35\% peak value at 400 nm at room temperature, guaranteed to be stable at LAr temperature. The main characteristics of the new R11065, high QE Hamamatsu PMT are reported in Fig.\ref{fig:hmmts+PMT}. The specific QE as a function of the incident wavelength, as reported by Hamamatsu, for one of the PMTs used in the series of tests reported in this work, is shown in Fig.\ref{fig:QE}.\\
A high Collection Efficiency of photoelectrons at the first dynode (CE, above 95\%) is obtained for cathode-to-first dynode voltage 
above 300 V. 
The voltage divider for the 12-stages dynode chain is custom made on a G10 printed circuit according to a Hamamatsu reference electrical scheme (AC coupling, 50 $\Omega$ anode termination to ground). All passive components  were selected for operation at cryogenic temperature.

\section{Light Detection in Liquid Argon}
\label{sec:light_lar}

\begin{figure}[h]
\vspace*{-0.5cm}
\begin{center}
\includegraphics*[width=9.0cm]{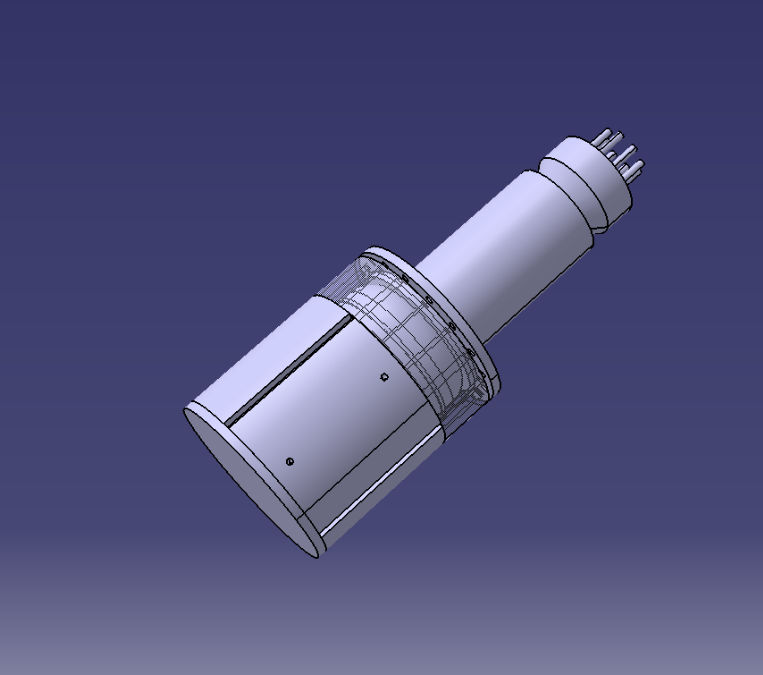}
\includegraphics*[width=4.45cm]{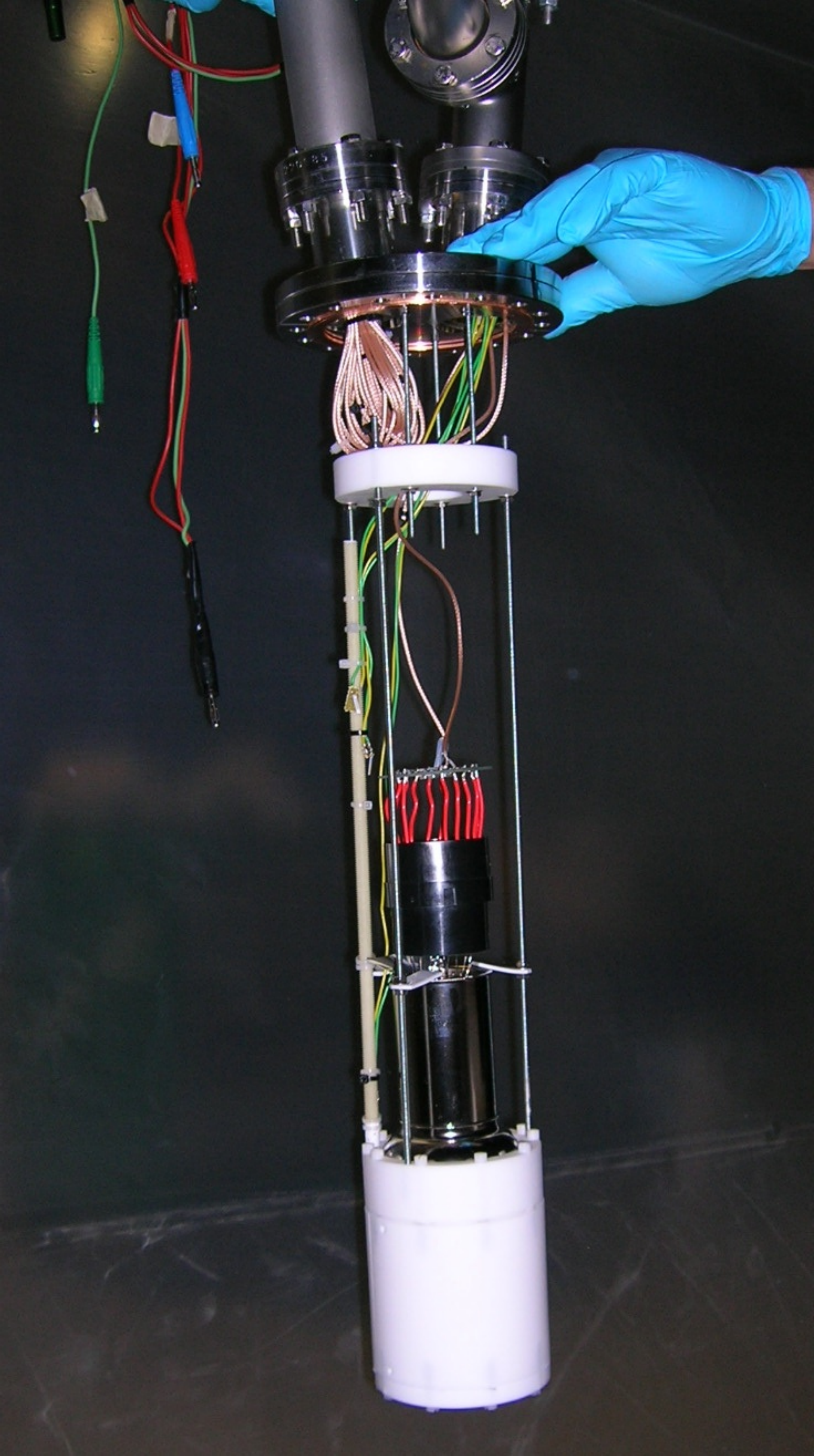}
  \caption{\textsf{\textit{3D isometric view and picture of the detector in use: PTFE cell and Hamamatsu PMT.}}}
\label{fig:LArCell}
\end{center}
\end{figure}

 Luminescence in liquid Argon is due to the radiative decay of low excited molecular states (Ar$^*_2$) formed in ionization events \cite{kubota,doke}.
 Photons are emitted with wavelength in the VUV range (around 127 nm) and exponentially  distributed in time with two (main) different time constants ($\tau_S\simeq$ 5 ns for the {\it fast} component and $\tau_T\simeq$ 1.3 $\mu$s for the {\it slow} component, corresponding to the decay of the excimers in {\it Singlet} and {\it Triplet} states respectively, as for example reported in \cite{heindl} from a recent dedicated measurement with time resolved techniques). \\
The transmittance of the PMT-window (Synthetic Silica glass) for VUV photons below 160 nm is null. Therefore, LAr VUV light has to be shifted to longer wavelengths. This is commonly accomplished by using efficient {\itshape wavelength shifter} materials ({\it wls}) such as {\itshape Tetraphenyl-Butadiene} (TPB) \cite{TPB_prop}.\\ 
The emission spectrum of TPB is peaked around the blue line at 440~nm and extends from 390 to 520~nm, where the transmittance of the glass window and the photocathode quantum efficiency (QE) of the PMTs are sufficiently high. In Fig.\ref{fig:QE} the TPB emission spectrum is shown superimposed to the R11065 quantum efficiency as a function of wavelength. \\
To effectively provide the LAr based detectors with a high light detection capability  and to make the detector response  homogeneous, all the internal surfaces delimiting the LAr sensitive volume have to be covered with a TPB layer. \\

In the tests reported here, the boundary surfaces of the detectors (side and bottom end) were completely surrounded with a highly reflecting layer coated by a thin TPB film (about 300 $\mu$g/cm$^2$) obtained by deposition with vacuum evaporation technique. The reflector layer (3M-VIKUITI ESR) was a polymeric, totally dielectric, multi-layer plastic mirror  with highest specular reflectivity ( 99\% ).\\
This configuration has been chosen to simultaneously optimize the down-conversion efficiency of the impinging VUV photons and the reflection efficiency of the blue-shifted photons. In this way, scintillation VUV photons from energy deposition in the LAr volume propagate inside the LAr volume and are wavelength-shifted into visible photons when hitting the TPB film on the surface boundaries.  The TPB film + reflector underlayer  has a high reflectivity to the visible
 photons (around 95$\%$) meaning that down-converted visible photons can be reflected (several times) from the  boundary surfaces\footnote{In other LAr detectors a thin TPB layer is usually deposited (embedded in a PoliStyrene matrix) onto the PMT window. In the tests reported here however the PMT window was left naked. This allows the highest visible photon collection at the PMT photocathode at the expense of losing the VUV light fraction directly impinging onto the PMT window.}, up to collection on the photocathode. \\

Effects of residual impurities in LAr on the scintillation light output may significantly degrade the detector performance \cite{warp_N2,warp_O2}. \emph{ Quenching} (i.e. non-radiative) processes in two-body collisions of impurity-molecules with Ar$^*_2$ excimer states (otherwise radiatively decaying with scintillation light emission) and \emph{Absorption} of the emitted VUV photons by photo-sensitive impurities can take place depending on the type of impurity and its concentration level. Light collection becomes affected by the presence of O$_2$ and H$_2$O molecules diluted at $\ge$ 100 ppb (part per billion) and nitrogen diluted at ppm (part per million) levels of concentration. The reduction of the O$_2$ and H$_2$O content by appropriate purification systems
(molecular sieves for water and Oxygen reactants) is definitively needed in Dark Matter detectors based on the collection of the LAr scintillation light.\\

\section{Single PMT test}
\label{sec:single_pmt}

The device for the first test with a single PMT was composed of a LAr cell (in PTFE, cylinder shaped with internal dimensions h=9.0 cm and $\phi$=8.4 cm)  with the 3" R11065 PMT, mounted face-down on the top side, viewing the  0.5 lt LAr volume inside the cell.  An isometric layout and a picture of the set-up are shown in Fig.\ref{fig:LArCell}.

The PTFE cell was housed in a long stainless steel cylindrical
 vessel. Its internal volume is about 5 lt and it contains, after filling, 
a total amount of at least 3.5 lt of LAr in order to have the PMT and its base fully immersed.  
The LAr active volume of the detector cell (about 0.5 lt) was optically independent but not partitioned from the rest of the LAr volume inside the vessel.
The vessel was deployed in a LAr bath of an open dewar, to keep the LAr internal volume at stable temperature.
The experimental set-up was assembled and operated at the \WArP Cryogenic Facility (LNGS - External Laboratory).\\
This  layout is very similar to that in use for a former set of experimental measurements reported in \cite{warp_N2,warp_O2}, to which we refer to for more details on the detector set-up.

The first test of the Hamamatsu PMT has been carried out in fall '09 and repeated again in early 2010 for verification of the results obtained. 
Each test run lasted about 10 days after detector activation.
The activation procedure consisted of a vacuum pumping phase of the vessel down to few $10^{-5}$ mbar. The residual gas composition measured with a mass spectrometer indicated H$_2$O as the main component, due to material outgassing from detector components. After immersion of the vessel in the LAr bath water outgassing was quickly halted due to the temperature dropping below freezing point. This was indicated by the residual pressure drop inside the chamber by more than one order of magnitude. At this stage the LAr filling procedure through an in-line set of filtering cartridges (Oxygen reactant and molecular sieve, Oxisorb and Zeolite) was started and smoothly completed in about one hour.

After a period left for thermalization of the PMT at LAr temperature (about one day) the bias voltage on the PMT was slowly raised up to working conditions.

The DAQ system was structured with the PMT anode current output directly transmitted through a LEMO cable (50 $\Omega$) to a fast {\it Waveform Recorder} ({\sf Acqiris, DP235 Dual-Channel PCI Digitizer Card}, up to 1 GS/s, 8 bit dynamic range). At each trigger the signal waveform is recorded with sampling time of 1 ns  over a full record length of 15 $\mu$s \\

\subsection{Data Analysis and Results}
\label{sec:SER}
\begin{figure}[htb]
\begin{center}
\includegraphics*[width=17.0cm, height=6.0cm]{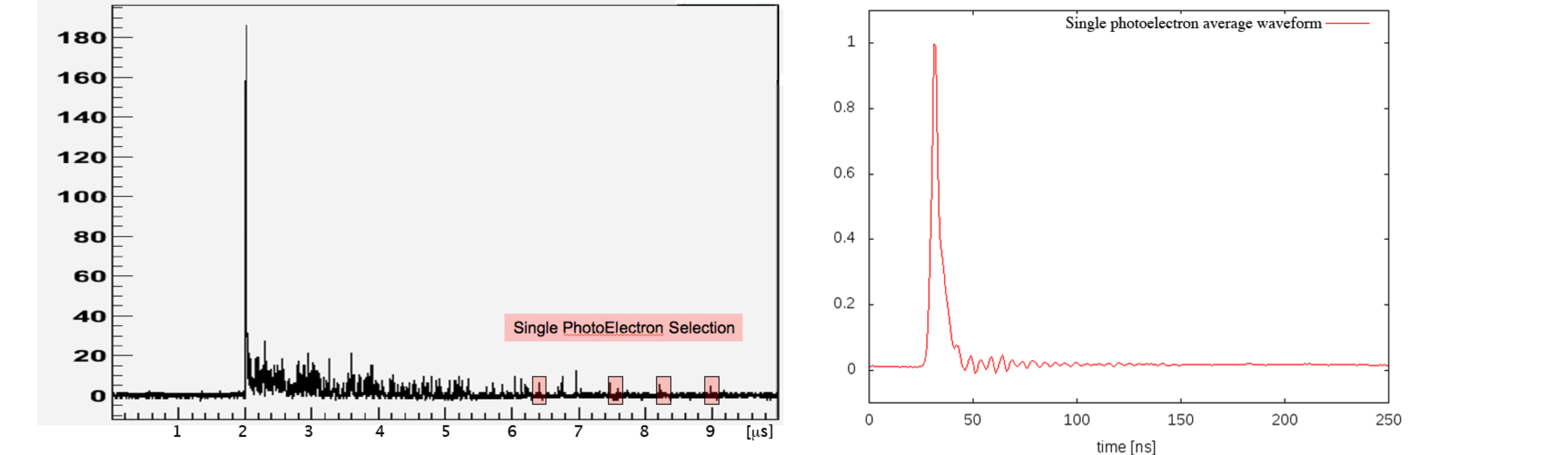}
 \caption{\textsf{\textit{Hamamatsu R11065 PMT test: [Left] example of recorded waveform ($^{241}$Am source). Single photo-electrons are identified (red shadow) in the waveform tail. [Right] SER averaged pulse.}}}
\label{fig:wfm-SER}
\end{center}
\end{figure}
\begin{figure}[htbp]
\begin{center}
\vspace{-0.2 cm}
\includegraphics[width=9.5cm]{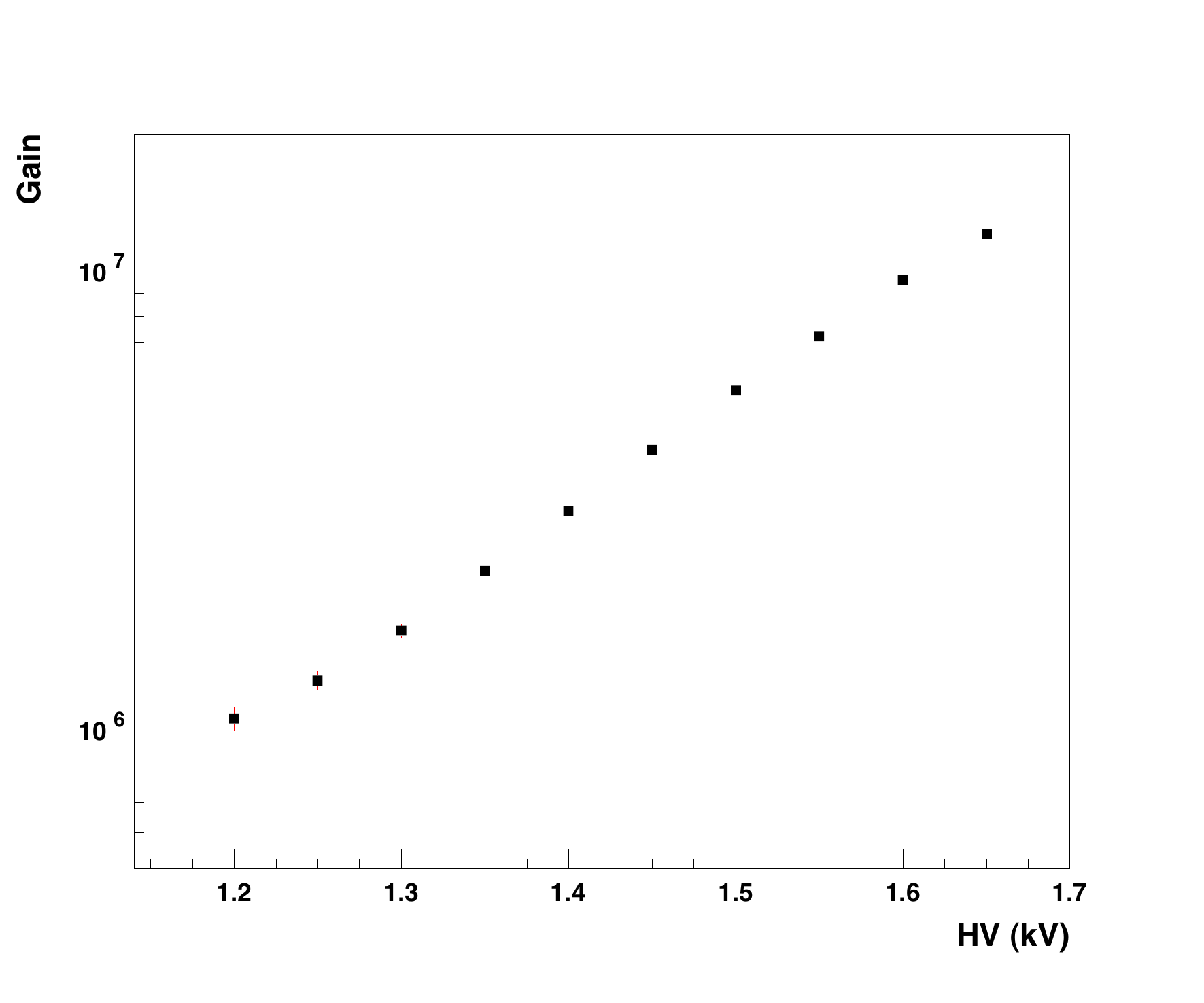}
 \vspace{-0.3 cm}
\caption{\textsf{\textit{Hamamatsu R11065 PMT test: Gain dependence on HV (measurement at LAr temperature).}}}
\label{fig:HMMTS-gain-test}
\end{center}
\end{figure}

During each source run (one per hour), single photo-electron (SER) pulses have been selected from out-of-trigger parts of the recorded waveforms (isolated peaks in the waveform tail, Fig.\ref{fig:wfm-SER} [Left]), in order to obtain photo-electron data needed for calibration.\\
Typically, the SER pulse is a narrow signal ($FWHM~\simeq~5$ ns) quickly returning to baseline (about 20 ns from  the onset). An averaged single photo-electron pulse is shown in Fig.\ref{fig:wfm-SER} [Right].\\
\begin{figure}[htb]
\begin{center}
\includegraphics*[width=7.cm, height=5.5cm]{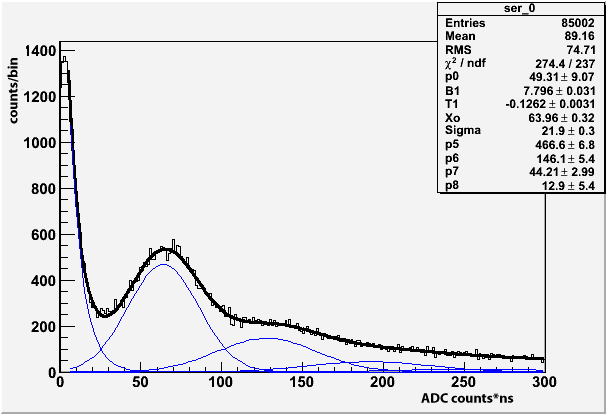}
 \caption{\textsf{\textit{Hamamatsu R11065 PMT test:  SER spectrum (PMT bias voltage +1400 V).  The fit superimposed is obtained by the sum of an exponentially falling function (dark count distribution) and three Gaussian functions for the single photo-electron (first peak; fit parameters $\mu=X_0$ and $\sigma$=Sigma) and multiple photo-electron distributions.}}}
\label{fig:ser_spectrum}
\end{center}
\end{figure}
The area under the selected peak (in {\it ADC$\cdot$ns} units, proportional to the SER charge) is evaluated by integration of the single photo-electron pulse after local baseline subtraction\footnote{The local baseline is evaluated in a 50 ns window starting 25 ns after the peak when no other peaks are observed inside the baseline window. If a second peak is found, it is merged with the previous one, the baseline window is set at 25 ns from the second peak and the check is performed again. This algorithm results in a charge spectrum that corresponds to the superposition of single and multiple photoelectron distributions.}.  SER charge spectra were obtained for each source run, all throughout the test period. A typical spectrum is reported in Fig.\ref{fig:ser_spectrum} (PMT bias voltage +1400 V): the Gaussian distribution around the first peak is well separated from the thermionic dark counts distribution and corresponds to the genuine single photo-electron mean amplitude and spread. The other broader peaks at higher charge values are associated to multiple, 2 or 3, photo-electron distributions.\\
The position of the first peak allows the monitoring of the gain setting\footnote{The gain $G$ of the PMT multiplication system corresponds to the output charge collected for a (unitary) input charge (single photo-electron). The output charge is estimated from the (first) peak of the SER spectrum after conversion from [ADC$\cdot$ns] units into charge units [p$C$] (k=50/2$^8$[mV/ADC]$\cdot$(1/R$_1$+1/R$_2$), with R$_1$=R$_2$=50$\Omega$ - parallel of the termination to ground of the voltage divider and of the transmission line to DAQ respectively.} (and gives the calibration constant per single photo-electron, useful for the corresponding gamma source spectrum analysis). \\ 
The gain dependence on the High Voltage applied to the PMT has been measured by changing the HV setting and measuring the corresponding position of the SER peak in the charge spectrum. The result is shown In Fig.\ref{fig:HMMTS-gain-test}. The nominal gain $G~= ~5\times 10^6$ (Hamamatsu data sheet) is achieved at a voltage supply of +1500 V. 

\begin{figure}[htbp]
\begin{center}
\vspace{-0.2 cm}
\includegraphics[width=7.2cm]{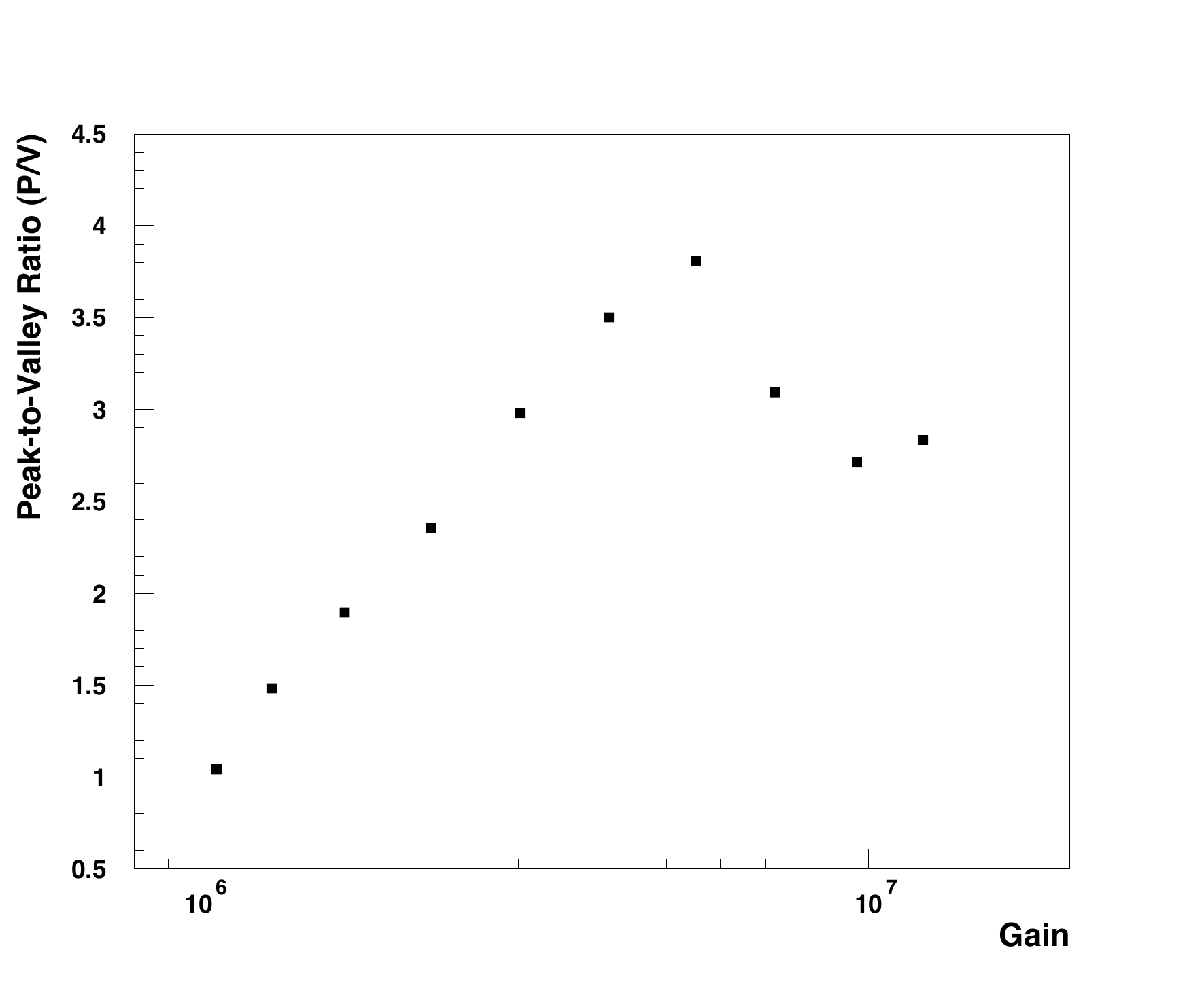}
\includegraphics[width=7.2cm]{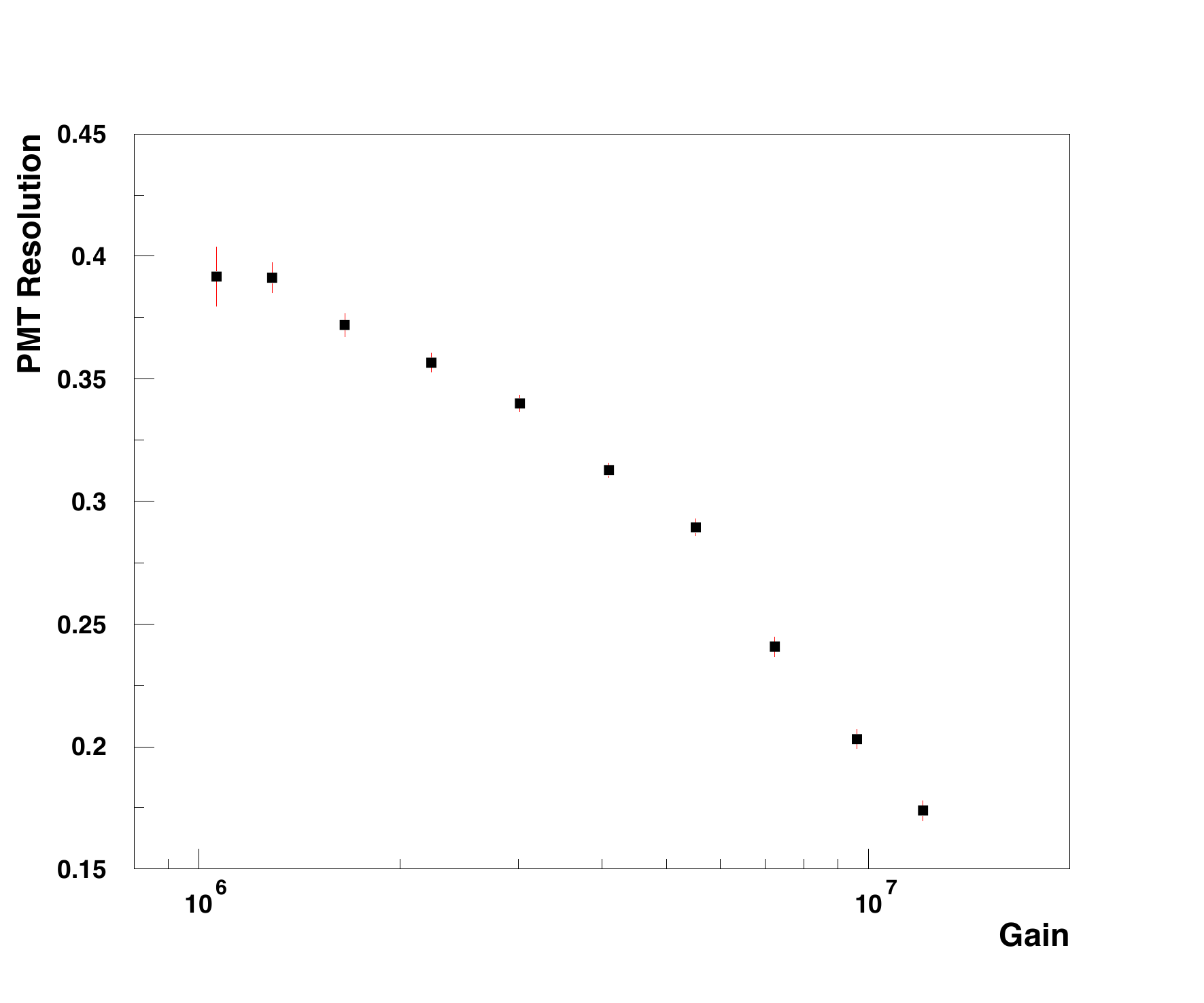}
 \vspace{-0.3 cm}
\caption{\textsf{\textit{Hamamatsu R11065 PMT test: SER Peak-toValley ratio determined at different gain values [Left]. SER peak resolution at different gain values [Right]. (Measurements at LAr temperature).}}}
\label{fig:ptov_vs_hv_newboard}
\end{center}
\end{figure}

 The SER Peak-to-Valley ratio, obtained by comparing the single photo-electron peak with the point where the single photo-electron distribution meets exponentially falling distribution attributed to the the dark counts, is usually taken as a figure of merit of the PMT performance.  In Fig.\ref{fig:ptov_vs_hv_newboard} [Left] the $P/V$ ratio determined at different gain values is shown: at the nominal gain it reaches the maximum value of $P/V\simeq ~3.7$. \\ 
The resolution of the SER peak, defined as the ratio $R=({\sigma}/{\mu})_{SER}$, is another important PMT characteristic to be measured at LAr temperature. This allows to infer the Excess Noise Factor (ENF) of the PMT, due to the fluctuations of the multiplication process in the dynode chain (ENF$=1+R^2$).\\
The PMT resolution at different gain values has been measured (Fig.\ref{fig:ptov_vs_hv_newboard} [Right]). At the nominal gain of $5\times 10^6$ the resolution of the R11065 PMT  is found to be $R~\simeq~28$\% and the Excess Noise Factor is ENF$~\simeq 1.08$.\\

\begin{figure}[htbp]
\begin{center}
\vspace{-0.2 cm}
\includegraphics[width=8.cm]{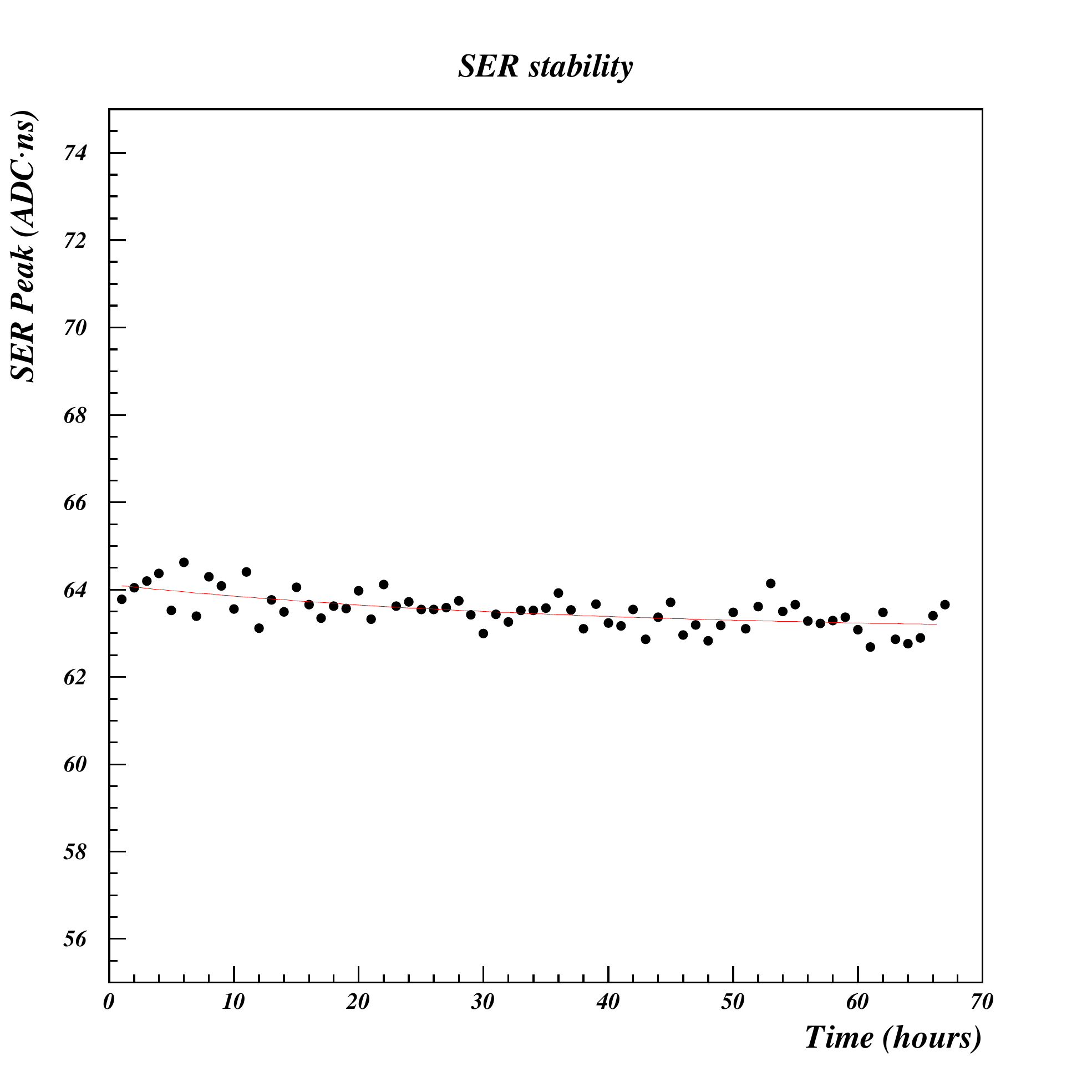}
 \vspace{-0.3 cm}
\caption{\textsf{\textit{Hamamatsu R11065 PMT test: SER peak stability in time (about three days) during the test run period.}}}
\label{fig:SER_stability}
\end{center}
\end{figure}

During the subsequent period of tests of the Hamamatsu tube a  gain setting lower ($G~= ~3.1\times 10^6$) than the nominal gain has been adopted (HV=+1400 V).  \\
Stability of the gain via the SER peak value has been monitored through the test period (about three days of data acquisition). The SER peak showed an almost stable behavior in time, as indicated by Fig.\ref{fig:SER_stability}, with a slightly decreasing exponential trend ($\tau\simeq 35$ h from the fit) attributed to residual effects of thermalization of the PMT at LAr temperature.  \\

The main objective of this run was the measurement of the detector light yield (LY) attainable with the use of this new PMT. The LY, commonly defined as the number of photo-electrons ({\it phel}) collected per unit of deposited energy in units of {\it phel}/keV, primarily depends on the PMT Quantum Efficiency but also upon several other factors, including the  detector geometry, the photo-cathodic coverage and the actual operating conditions of the experimental set-up. Among these last the LAr purity and the TPB wavelength-shifting efficiency  may play a significant role. Therefore, the evaluation of the experimentally determined LY needs to be compared with an a priori calculated reference expectation value  (e.g by means of MonteCarlo simulations).\\ 

 The LY measurement was performed by exposing the LAr cell viewed by the Hamamatsu PMT to a $^{241}$Am monochromatic $\gamma$-source  with emmission at 59.54 keV to obtain a reference energy deposit in LAr. The source was located inside a collimator holder positioned outside the cell in a fixed position. The trigger rate has been monitored  giving a stable  value of $\sim 180$ Hz during the source runs. Data acquisition runs with the source have been alternated with blank runs (background from ambient radiation). \\
Gamma rays from the $^{241}$Am source  
induce photo-electric interactions in the LAr cell active volume\footnote{Considering the LAr mass attenuation at this energy ($\simeq$0.5 cm$^{-1}$) a good fraction of the LAr volume of the cell is exposed to  photo-electric interactions (without locus dependent bias).}, with electron emission in the {\it mip} range. 

Scintillation light following the electron energy deposition, after down-conversion and reflections 
at the active volume boundaries, is collected at the PMT photo-cathode and the signal waveform is recorded. \\
\begin{figure}[htbp]
\begin{center}
\includegraphics[width=10.cm]{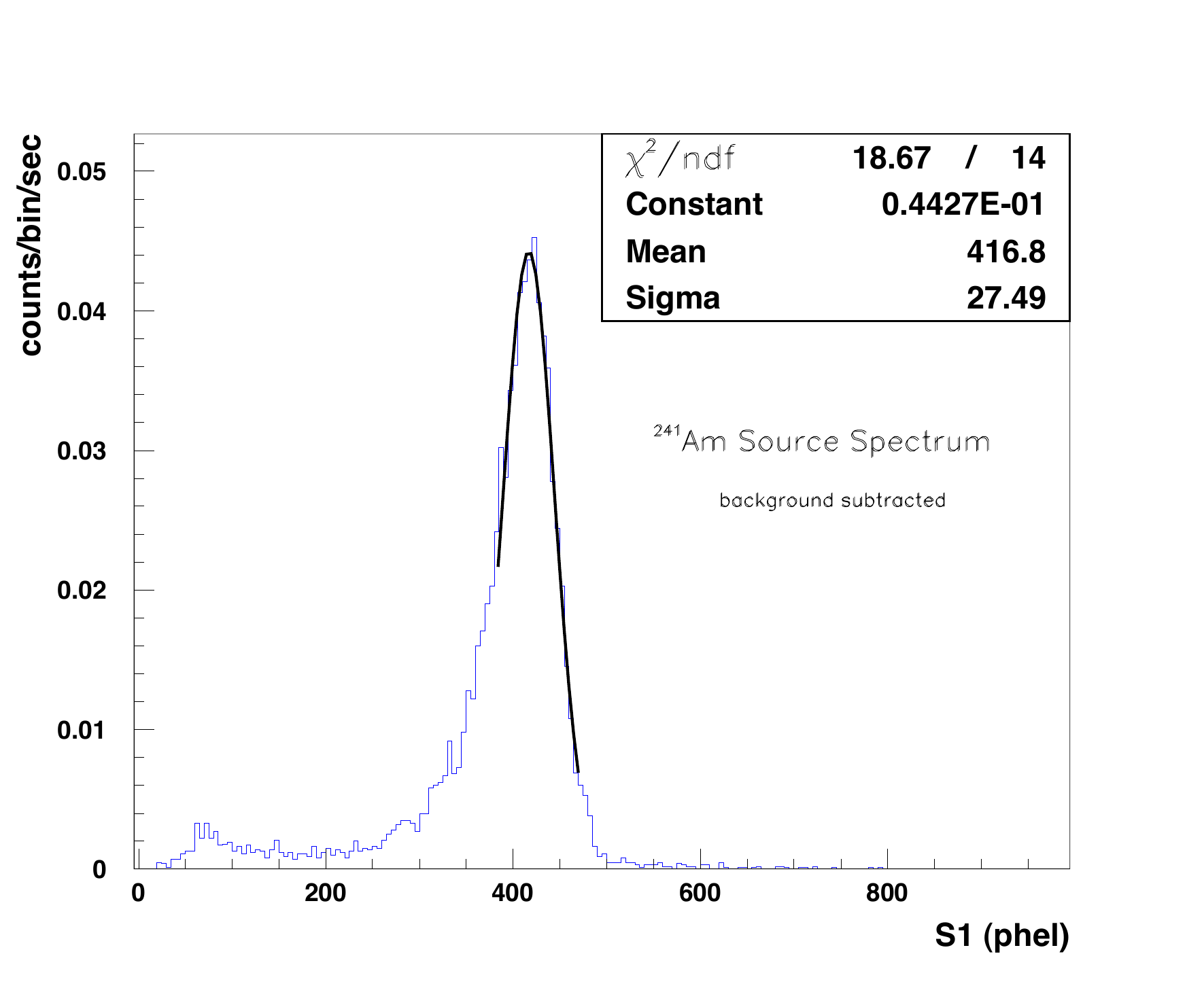}
\caption{\textsf{\textit{Hamamatsu R11065 PMT test: $^{241}$Am source spectrum (blank spectrum subtracted). The relative energy resolution at the peak energy is $\frac{\sigma_E}{E} \simeq$ 7\%.}}}
\label{fig:241Am_spectrum_1}
\end{center}
\end{figure}

By waveform integration, after local baseline evaluation and subtraction, the event signal amplitude $S1$ was obtained in ADC. It was normalized with the values obtained by fitting the SER spectra for each run giving its value in {\it phel} units.
The pulse amplitude is proportional to the electron energy deposited in the LAr cell.  \\ 
Standard cuts have been applied to remove low energy events (E$_{min} = 20~phel$), high energy ADC saturated events, pile-up and out of time events. 
Pulse amplitude spectra have been thus obtained for each source run. 
\\





A  $^{241}$Am spectrum from one of the source runs collected 
 is shown in Fig.\ref{fig:241Am_spectrum_1}. 
As reported in the figure, the fit of the full absorption peak is found at 416 {\it phel}. Therefore, assuming full deposition of the 59.54 keV, the Light Yield of the detector can be evaluated as:
\begin{equation}
LY~=~7.0~\frac{phel}{{\rm keV}}~\pm~5\%
\end{equation}
\begin{figure}[htbp]
\begin{center}
\vspace{-0.4 cm}
\includegraphics[width=11.cm]{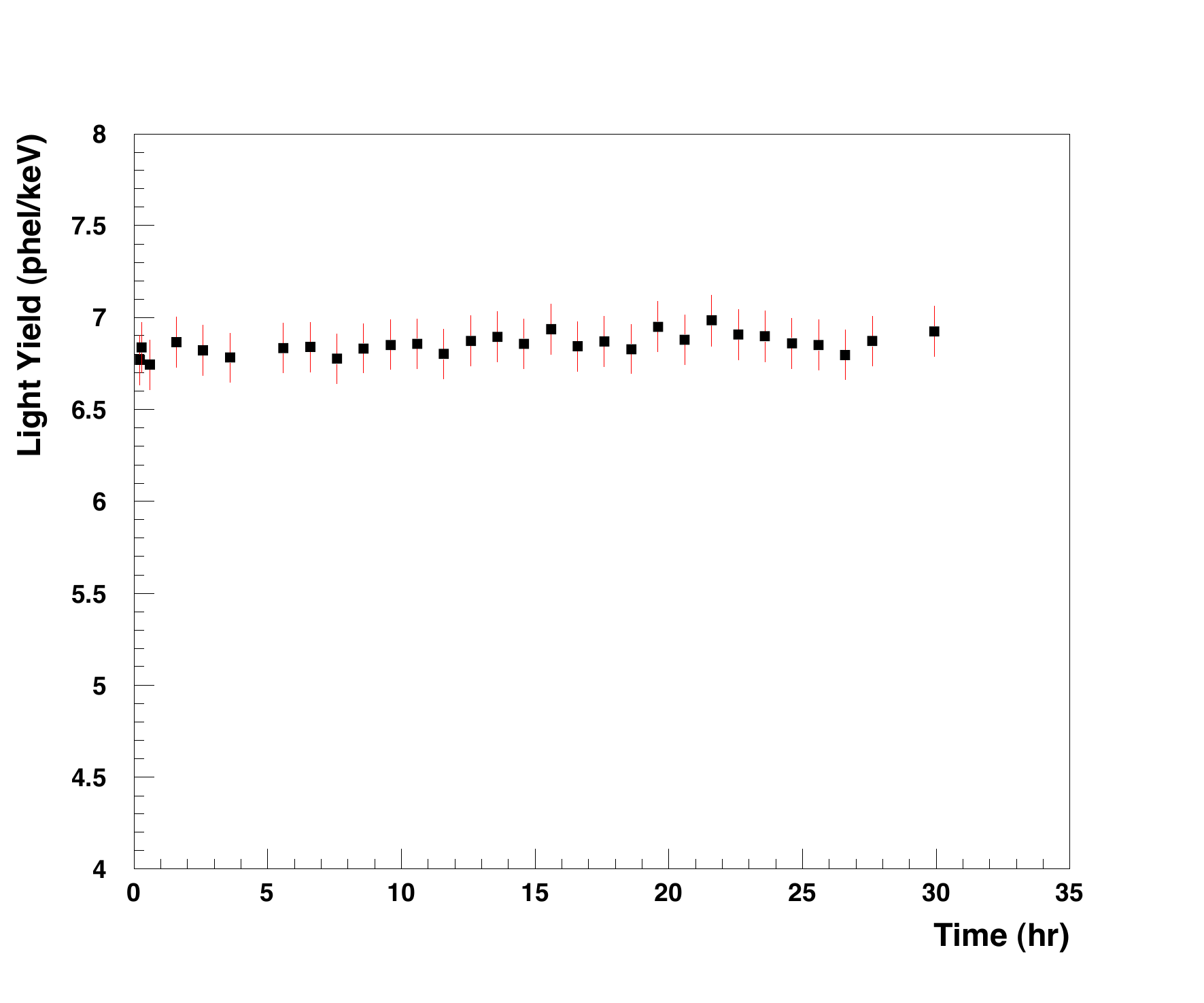}
 \vspace{-0.3 cm}
\caption{\textsf{\textit{Hamamatsu R11065 PMT test: LY stability in time.}}}
\label{fig:LY-stability}
\end{center}
\end{figure}

The statistical error from the fit is negligible and
the systematic error associated to the LY measurements is evaluated from the dispersion of the photo-peak values in the collected $^{241}$Am spectra (4 to 5 \%, background subtracted),  from uncertainties associated to the calibration ($\simeq$ 2\%, SER peak determination) and also from the systematics of the off-line analysis due to the choice of internal parameters  (evaluated by varying them within 2 to 3\%). An overall systematic error around 5\% is estimated.

This remarkable result\footnote{As a term of reference, with a similar experimental set-up (0.5 lt cell) equipped with another  type of PMT (Electron Tube, 3"  ETL Mod. D750 - nominal QE=20\% at 420 nm \cite{ETL-D750}), the measured LY yielded a maximum value around 2.4 {\it phel}/keV.} is fully compatible with the MonteCarlo expectation value
based on standard assumptions on the optical properties of the reflector+TPB coating and on the LAr purity level estimated during the detector operations. It confirms the very good performance of the R11065 Hamamatsu PMT at LAr temperature in line with expectations from its nominal Quantum Efficiency.\\


The stability in time of the LY  has been monitored by applying the analysis to a large set of runs collected during the test period.
The results are shown in Fig.\ref{fig:LY-stability}. All LY values lie  within $\pm 1.5 \%$ around a mean value of 6.9 {\it phel}/keV.\\

At the end of the test run the chamber was smoothly emptied and left in GAr atmosphere. After few weeks the detector has been refilled with a new batch of LAr, without any other change of the set-up (i.e. same PMTs, same TPB coated reflector surfaces).
The LY measurements from this additional test fully confirmed  the result reported above.

\section{ Direct Comparison of the R11065 with an ETL PMT }
\label{sec:2pmt}
\begin{figure}[h]
\begin{center}
\includegraphics[height=9.cm]{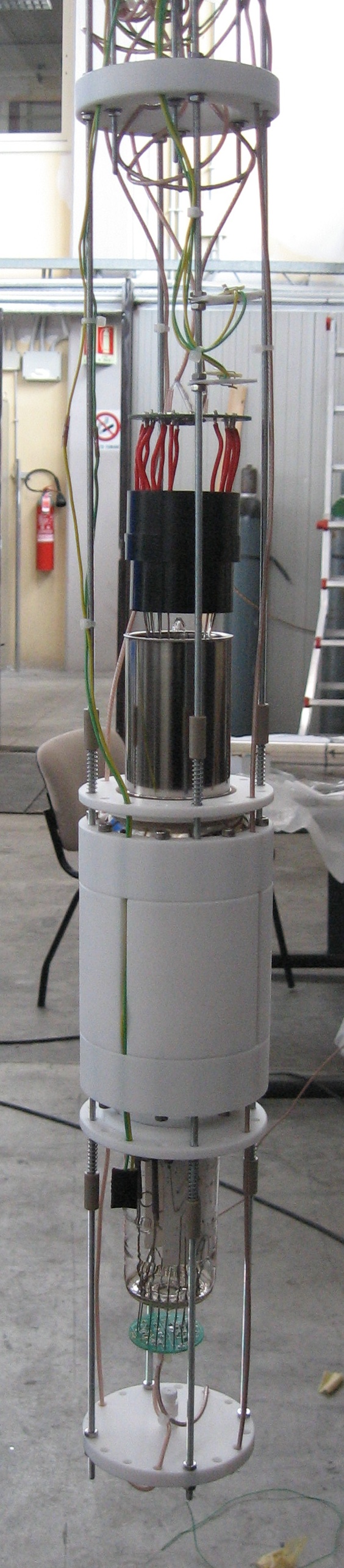}
\includegraphics[height=9.cm]{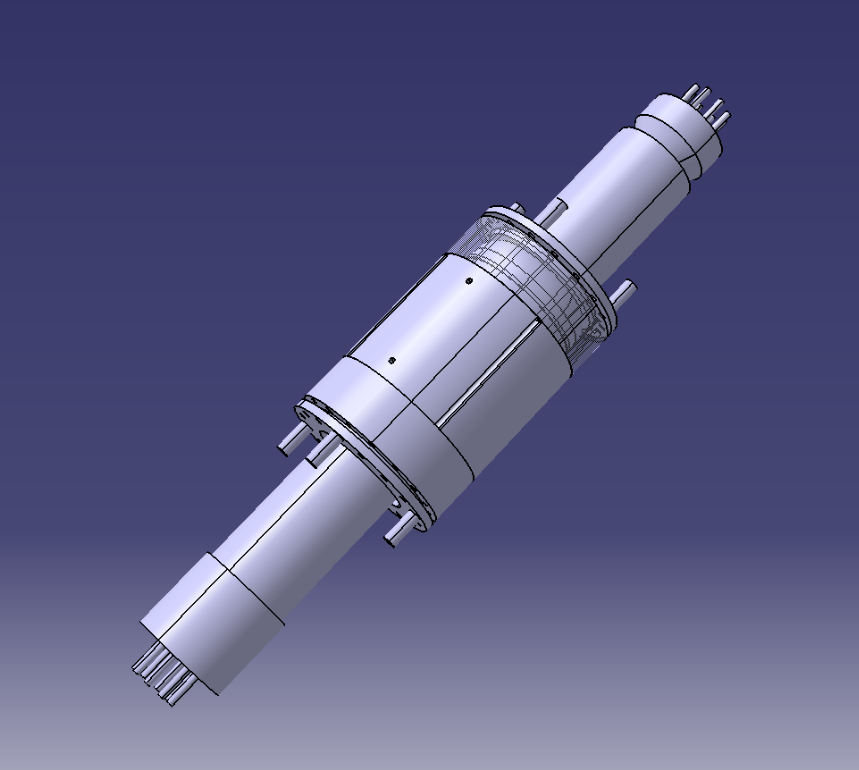}
\caption{\textsf{\textit{Layout of the TWO-PMT detector: on top the Hamamatsu PMT (face-down), at the bottom the ETL PMT face-up and in between the LAr cell in PTFE. The detector is fully immersed in LAr during operation.}}}
\label{fig:IMG_0097-2pmt}
\end{center}
\end{figure}

 The results reported above showed the excellent performance of the new Hamamatsu R11065, which seemed to be superior with regard to PMTs previously used by the WArP collaboration \cite{warp_100, warp_2.3}. However, the improvement in performance could, in theory, be an effect of differences in detector geometry and operating conditions.  In order to fully decouple from these effects a second dedicated test has been envisaged (mid 2010) for a direct comparative test of two types of PMTs: one 3" HQE Hamamatsu R11065 (used in the single PMT test) and one 3" ETL - D750 (pre-production series of the PMT type \cite{ETL-D750} adopted in the \WArP-100 experiment \cite{warp_100}). \\  
A picture of the detector chamber used in this test is shown in Fig.\ref{fig:IMG_0097-2pmt}.  It is made of a PTFE cell, about 0.4 lt of internal volume (h=8.0 cm and $\phi$=7.6 cm) for LAr, 
lined with a TPB coated reflector layer on the lateral wall, analogous to the single PMT test chamber. The PMTs were installed at both ends of the cell -the HQE Hamamatsu face down on top and the ETL face up at the bottom, both with naked windows not covered in wavelength shifter. 

\begin{figure}[h]
\begin{center}
\includegraphics[height=9.cm]{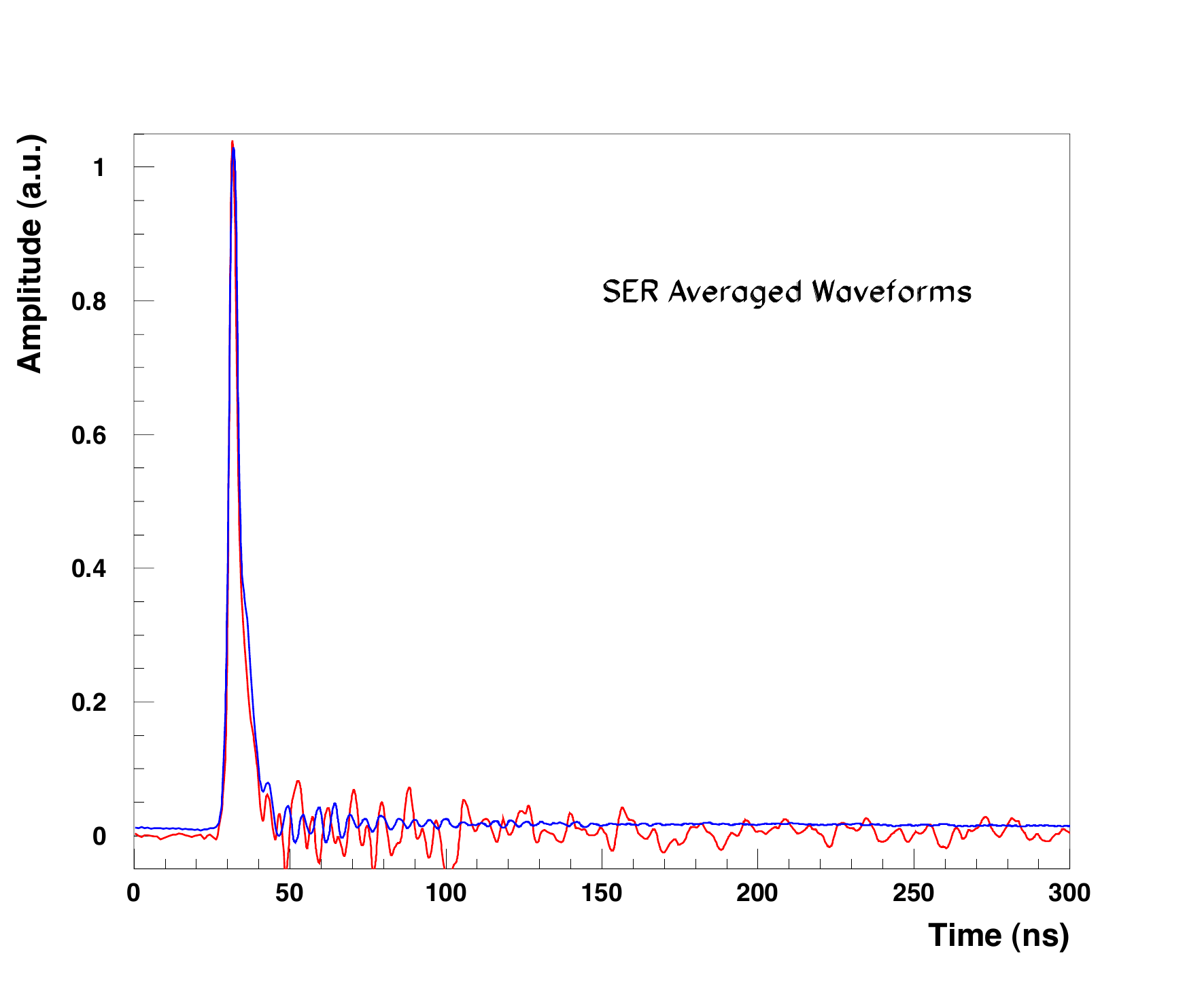}
\caption{\textsf{\textit{Hamamatsu vs ETL Test with the 0.5 lt LAr cell.  Single {\it phel} (averaged) Pulse Shape: Hamamatsu PMT [blue],  ETL PMT [red].}}}
\label{fig:avg-SER-pulse}
\end{center}
\end{figure}

The read-out, data treatment and the off-line analysis codes were the same as in the single PMT test (Sec.\ref{sec:single_pmt}). \\
Similarly, the single photo-electron pulses were identified and selected for each of the two channels leading to the single photo-electron averaged pulse for each PMT, Fig.\ref{fig:avg-SER-pulse}. The two PMT pulse shapes are quite similar to each other. Returning to baseline is however faster for the Hamamatsu SER pulse, without presence of wiggles and oscillations exhibited by the ETL PMT.\\

\begin{figure}[h]
\begin{center}
\includegraphics[height=11.cm]{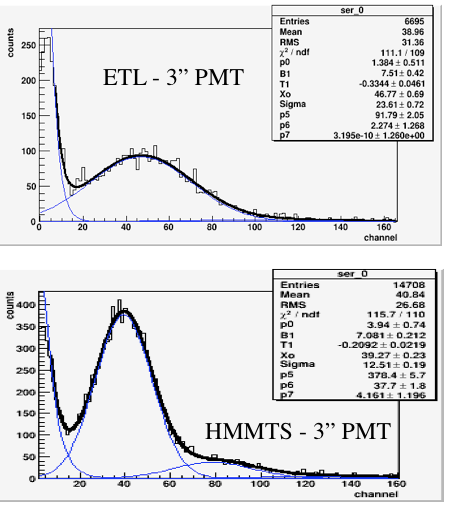}
\caption{\textsf{\textit{Hamamatsu vs ETL Test with the 0.5 lt LAr cell. SER Spectra: ETL PMT [top],
Hamamatsu PMT [bottom] (first peak; fit parameters $\mu=X_0$ and $\sigma$=Sigma).}}}
\label{fig:SER-ETL-HMMTS}
\end{center}
\end{figure}

SER spectra were obtained from each run and one example of typical SER spectra from the two PMT's is shown in Fig.\ref{fig:SER-ETL-HMMTS}. 
The positions of each peak were later used for gain calculation and calibration of respective acquired waveforms. 
\begin{figure}[h]
\begin{center}
\includegraphics[height=9.cm]{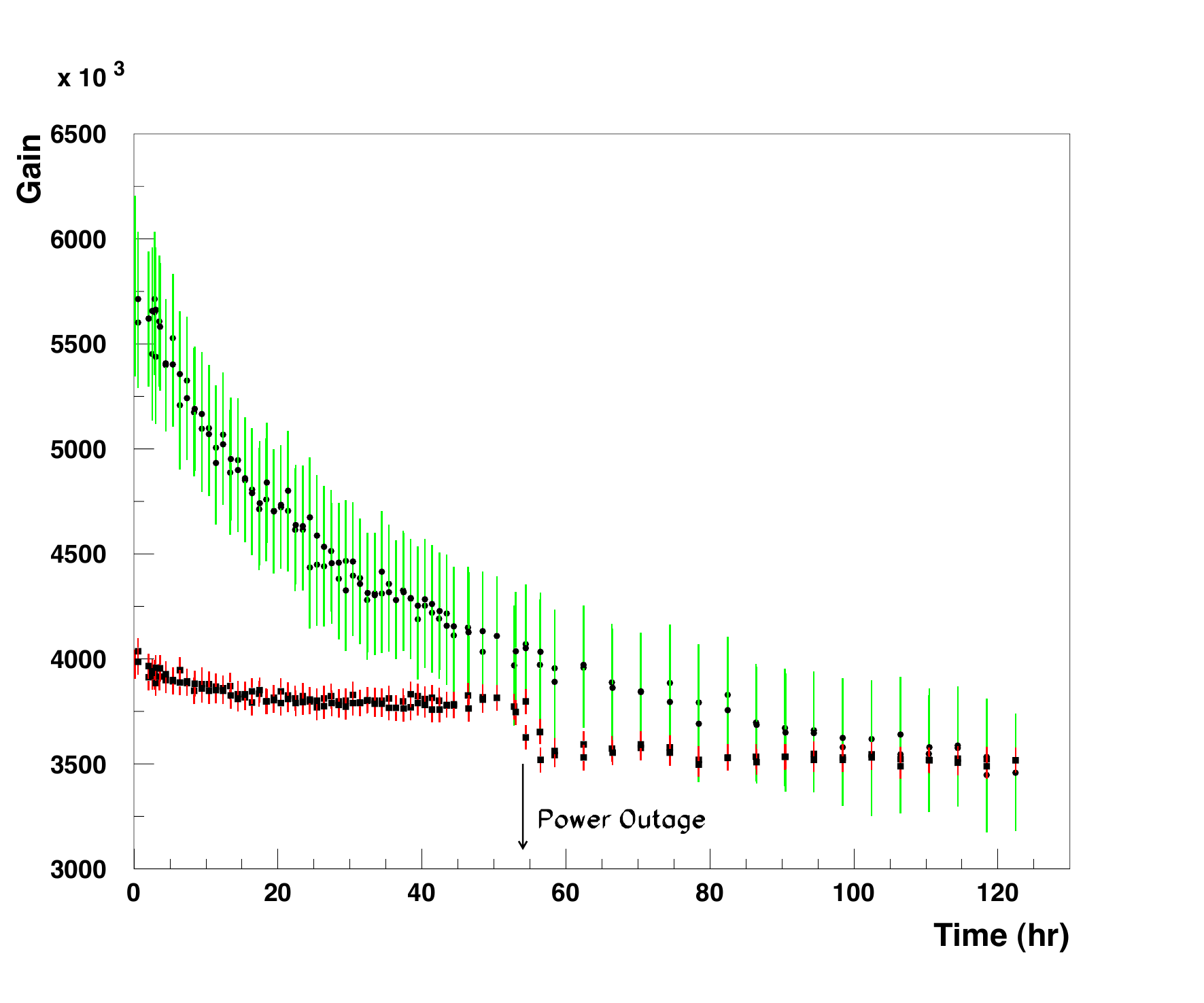}
\vspace{-0.3 cm}
\caption{\textsf{\textit{Hamamatsu vs ETL Test with the 0.5 lt LAr cell: gain variation in time (ETL PMT  [dots with green error bar],  Hamamatsu PMT [squares with red error bar]).}}}
\label{fig:gain-comparison}
\end{center}
\end{figure} 
\begin{table}[htbp] 
\begin{center} 
\caption{\textsf{\textit{Characteristic features of the Hamamatsu and ETL PMTs (Peak-to-Valley ratio and SER resolution) at fixed gain $G \simeq 3.7\times 10^6$.}}}
\vspace*{0.2cm}
\begin{tabular}{llcr} 
\hline\hline 
                 &    Peak-to-Valley ratio       & SER resolution     \\
\hline
Hamamatsu R11065       &  ~~~~~~~~3.5    &  	  32~\%	\\
ETL D750 (pre-series)    &  ~~~~~~~~1.9    &  	  50~\%       	\\
\hline\hline
\end{tabular} 
\label{tab:pv-res}
\end{center}
\end{table} 
After the LAr filling of the chamber one day was left for PMT thermalization. Subsequently, the Gain Stability in time has been monitored over about one week of run, Fig.\ref{fig:gain-comparison}. The initial gain of the ETL PMT has been set at a higher value, expecting a large decrease over the first days of operation (as observed in previous tests).  \\
Indeed, the gain of the ETL PMT showed a (quite steep) decreasing trend over the period of the measurements. The gain of the Hamamatsu PMT instead exhibited a slight decrease over the first day after activation and then stabilized to a constant value. Unfortunately, during the second day of operation an unexpected power outage occurred at the experimental site (\WArP cryogenic facility - LNGS). The PMT HV power supplies were powered using an UPS but the effect of the power outage is visible on the R11065 PMT as a sudden drop of gain, which was slowly recovered over the following days of running.\\
The cause of the much steeper gain loss and much longer stabilization time of the ETL tube is not clear. 
A gain of 3.7$\times 10^6$ has been set for both PMTs for the subsequent set of source runs.  At this gain value the characteristic features of the two PMTs (Peak-to-Valley ratio and SER resolution) are reported in Tab.\ref{tab:pv-res} for comparison.\\

\begin{figure}[h]
\begin{center}
\includegraphics[height=11.cm]{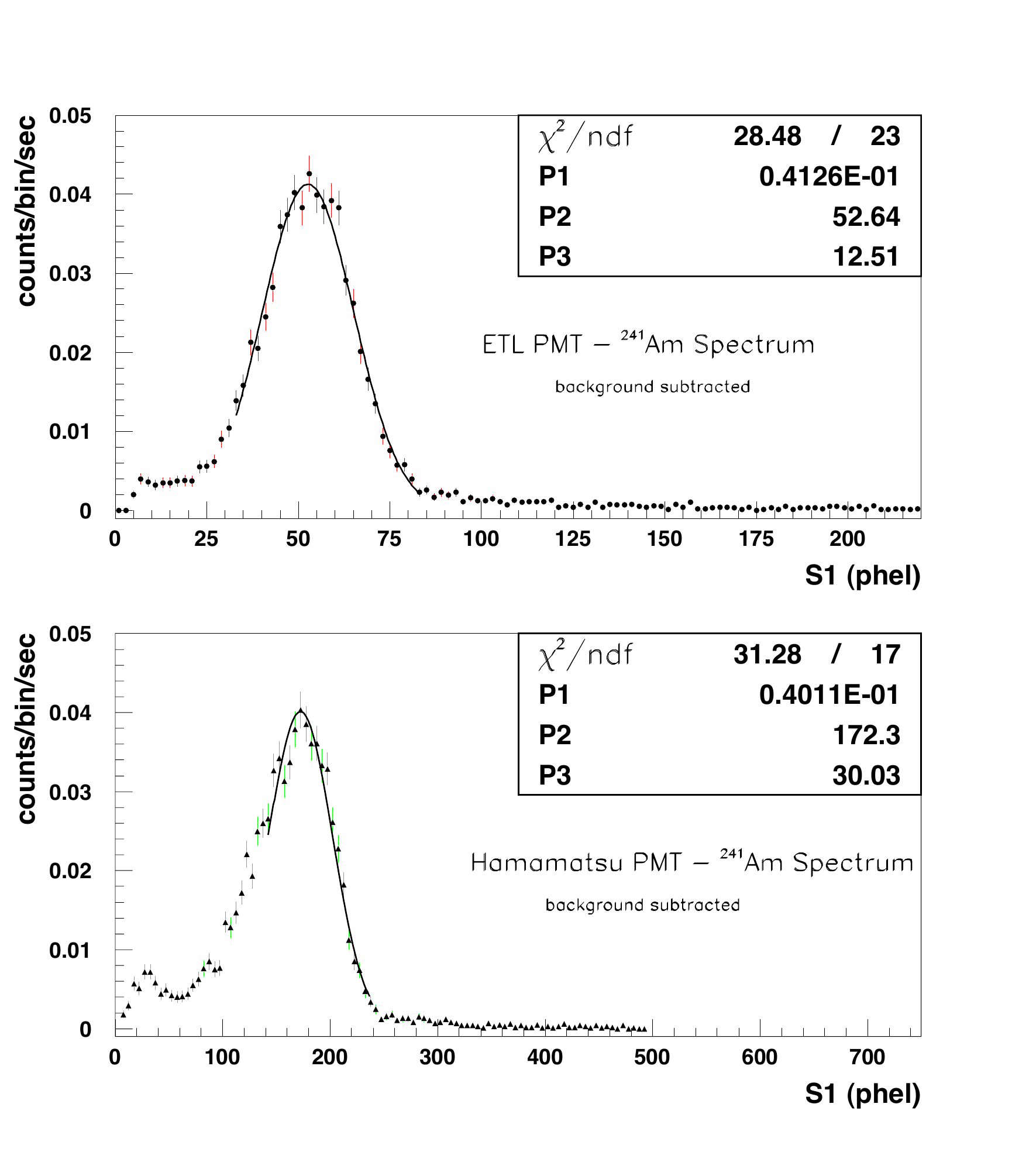}
\caption{\textsf{\textit{Hamamatsu vs ETL Test with the 0.5 lt LAr cell: $^{241}$Am full absorption peaks for the two PMTs ([Top] ETL PMT and [Bottom] Hamamatsu PMT - background subtracted). The peak values are found at $S1_{Ham}\simeq$ 172 {\it phel} and at $S1_{ETL}\simeq$ 52 {\it phel}. 
The relative energy resolution values at the peak energy are $(\frac{\sigma_E}{E})_{Ham}\simeq$ 17\% and $(\frac{\sigma_E}{E})_{ETL}\simeq$ 24\%.}}}
\label{fig:two_peaks}
\end{center}
\end{figure} 
\begin{figure}[h]
\begin{center}
\includegraphics[height=6.0cm]{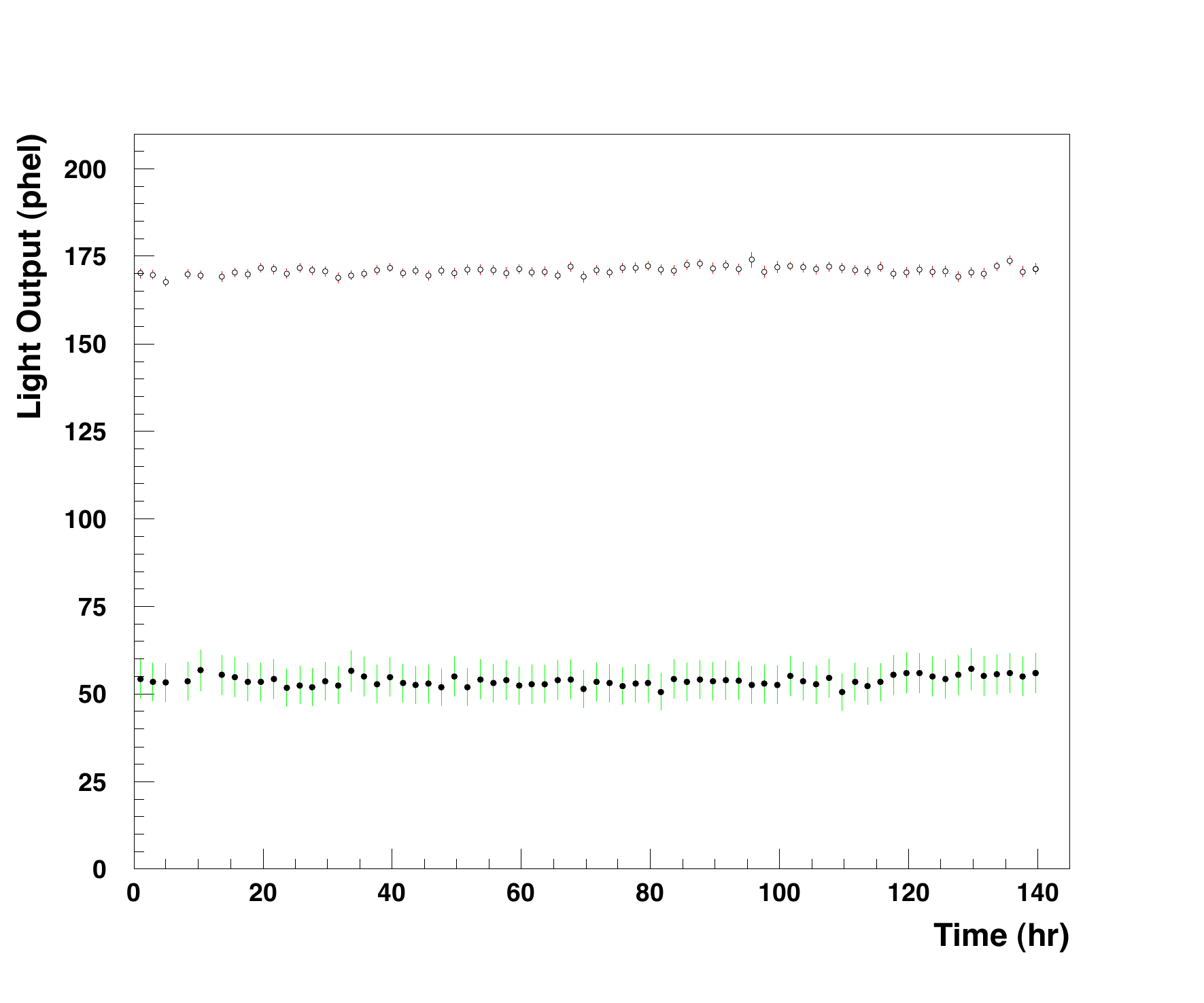}
\includegraphics[height=6.0cm]{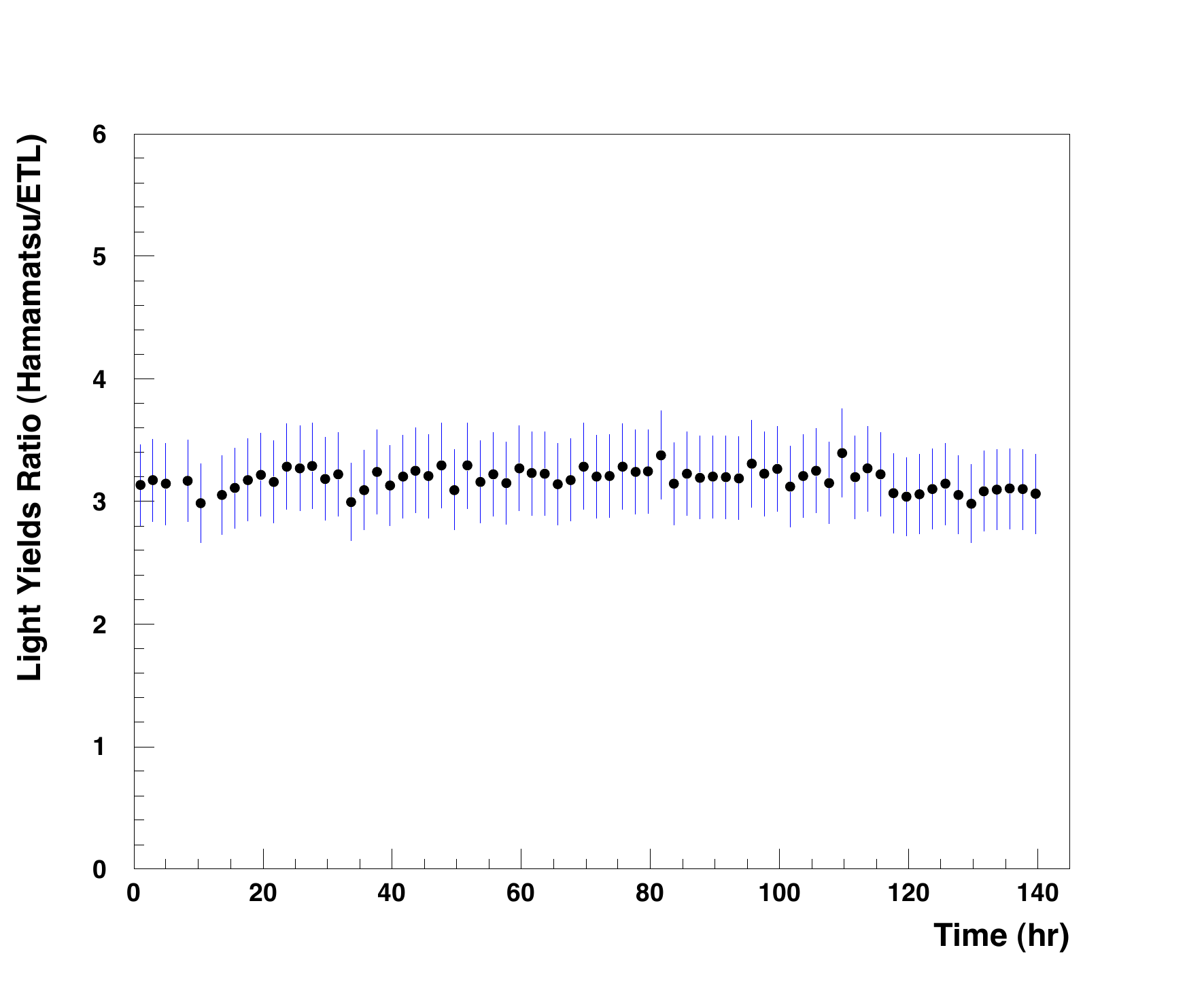}
\caption{\textsf{\textit{Hamamatsu vs ETL Test with the 0.5 lt LAr cell: [Left]  Mean Light Output (in [{\it phel}] units) - Hamamatsu 3'' PMT [dots],  ETL 3'' PMT [dots with green error bars, bottom]. [Right] Hamamatsu-to-ETL Light Yield ratio.}}}
\label{fig:LY-2pmt}
\end{center}
\end{figure} 
The main objective of the two-PMT direct comparison test was the determination of the individual light yield value for each PMT when operating under strictly identical conditions - i.e.  viewing the same LAr volume. This allowed decoupling from such variables as the purity of the liquid and the wavelength-shifting efficiency of the TPB film deposited onto the boundary reflective surface of the detector.\\
The light yield of the two PMTs has been determined by exposure to the $^{241}$Am gamma-source. As in the case of the Single PMT test, data acquisition runs with the source have been alternated with blank runs (background from ambient radiation). The signal amplitude for each event was obtained by integration and normalization to the respective SER position. \\
Pulse amplitude spectra for both PMTs have been thus obtained for each source run and used, by performing a gaussian fit, to determine the full absorption peak (typical mean value $\simeq$172 {\it phel} of light output for the Hamamatsu PMT and $\simeq$ 52 {\it phel} for the ETL PMT, Fig.\ref{fig:two_peaks}).  In Fig.\ref{fig:LY-2pmt} [Left] the mean light output values for the recorded runs, over a two-week period of operation, is shown. These values were then normalized to the source energy to obtain the Light Yield for each PMT.\\
A direct comparison between the two PMTs is the Hamamatsu-to-ETL LY response ratio which was found to be in the {\bf 3~:~1 range} (stable during the operation period), as shown in Fig.\ref{fig:LY-2pmt} [Right]. \\

Due to the configuration of the test set-up in use, the LY ratio depends only on the Global Efficiency	ratio of the two PMTs defined as $GE~=~QE\times CE$, where {\it QE} is the photocathode Quantum Efficiency and {\it CE}  the photoelectron Collection Efficiency at the first dynode.
The relative {\it QE} of the two PMTs has been measured in a facility at CERN. Its average value over the emission 
spectrum of the TPB results to be of the order of 2.7. Taking into account the {\it CE}: around $90-95\%$ for the 
Hamamatsu PMT and around $75-80\%$ for the ETL, the {\it GE} ratio is found to be in the region of $3.4 - 3.0$, well 
compatible with the measured LY ratio. 
 
\section{Four-PMT test}
\label{sec:4pmt}
The scaling-up of detector mass and complexity (e.g. increased number of PMTs) without loss of detector performance (lower LY) is in general not a trivial task and this is especially true for noble liquid detectors. A third experimental test has been thus performed to check if it is possible to repeat the obtained Light Yields in a detector about ten times bigger in volume compared to the one used with the first test reported above (Sec.\ref{sec:single_pmt}). Monitoring the stability of the system over an extended run period was the other main goal of this test.\\

The PTFE mechanical structure of the detector was taken from the \WArP-2.3 l prototype developed for a former set of experimental measurements \cite{warp_2.3} (we refer to it for more details on the detector set-up). The chamber was equipped with {\it four} R11065 HQE Hamamatsu PMTs. \\
The internal volume of the chamber is made of a cylindrical section ($\phi$ = 18.4 cm and h=9.5) superimposed to a slightly conical shaped section ($\phi_{top}$ = 17.6 cm, $\phi_{bot}$ = 16.0 cm  and h=7.5 cm), corresponding to about 4.3 lt  of total volume when completely filled up with LAr. 
The four PMTs were mounted face-down on the top side of the volume.\\
\begin{figure}[h]
\begin{center}
\includegraphics[height=10.cm]{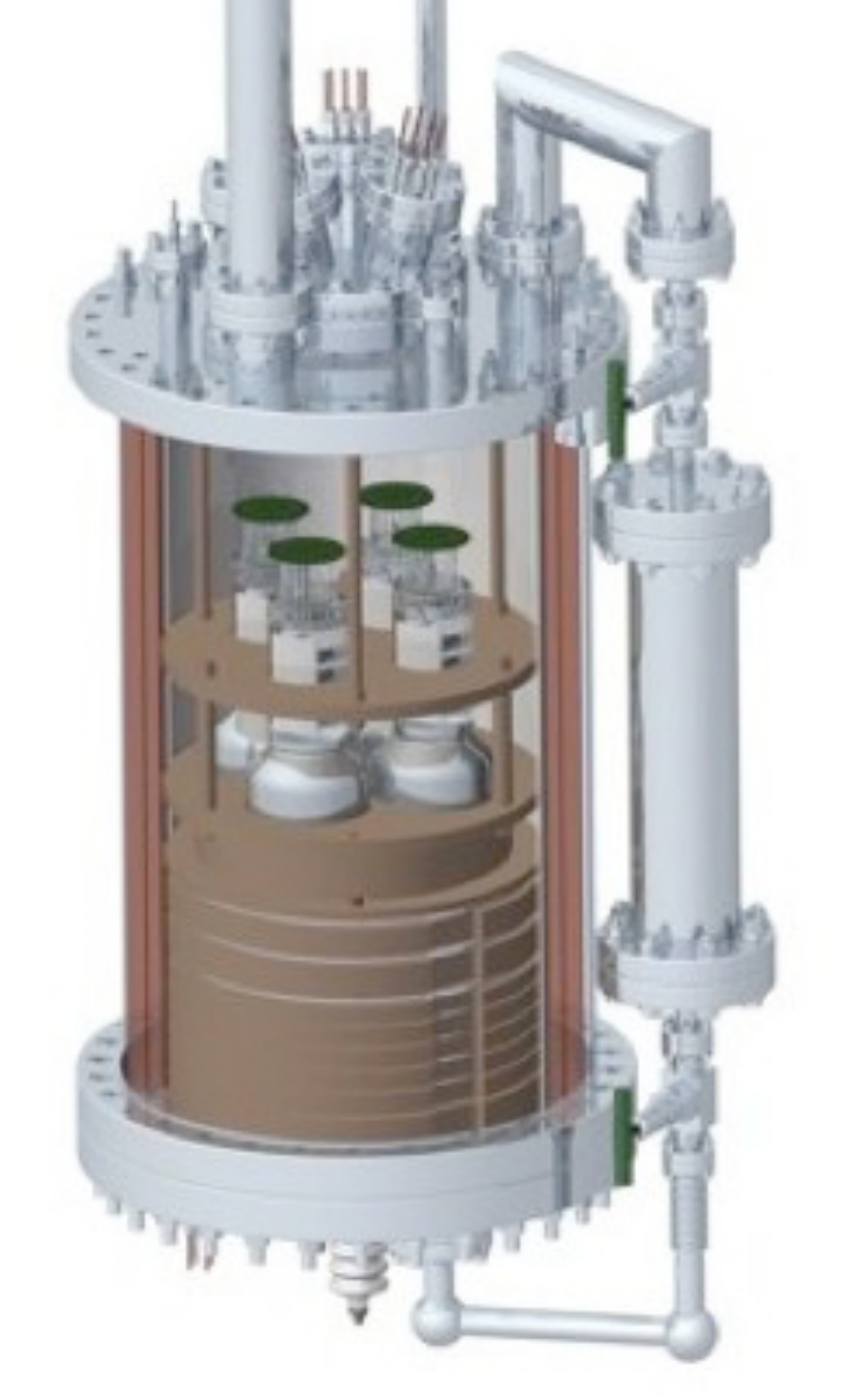}
\caption{\textsf{\textit\textbf{{Layout of the vessel and inner chamber. In the present test the LAr level inside the vessel is above the 
PMT bases (i.e. the detector is operated in single liquid-phase).}}}}
\label{fig:warp2_3l}
\end{center}
\end{figure} 
The boundary surfaces (lateral and bottom) were prepared as in the previous tests, similarly, the PMT glass windows were left naked. The photo-cathodic surface was about 12\% of the total boundary surface ($\sim$equivalent to the coverage of the "single PMT" test set-up). 
The detector was housed in a low-radioactivity stainless steel vessel, filled with purified LAr and immersed in a LAr bath of an open cryostat.\\
In Fig.\ref{fig:warp2_3l} a picture of the detector set-up is shown. \\ 
The 4 PMT anode signals were directly digitized by two Acqiris Boards (Mod. U 1080 A, 2-chs. each with 8-bit dynamic range and 1GS/s) at 1 ns sampling time over 15 $\mu$s time interval. This corresponds to the read-out chain currently implemented in the \WArP-100 experiment. DAQ and off-line codes were the same as for the previous tests reported above.\\

Before the detector assembly, all the PTFE mechanical components of the chamber  were baked in a vacuum oven (at around 80$^o$ C) for \textit{four weeks}. After assembly and mounting inside the vessel, the experimental set-up went through a first vacuum pumping phase (4 days) down to $10^{-5}$ mbar, followed by a warm GAr flushing phase (24 h) and again by a vacuum pumping phase (2 days, back to $10^{-5}$ mbar). Few hours after the immersion of the vessel in a 
LAr bath 
 the filling procedure through an in-line set of filtering cartridges (Oxygen reactant and molecular sieve) was started and completed in about three hours.\\
The vessel was completely filled up to full immersion of the PMTs and their bases in LAr. Therefore, the results reported below refer to measurements in single (liquid) phase at null electric field.   \\
Opposite to the previous tests, the PMTs were operated in DC coupling\footnote{The presence of the Voltage Bias decoupling capacitor needed for the AC coupling of the PMT introduces an unavoidable (though small) undershoot in the collected waveform shape around 10 $\mu$s after the signal onset, leading to an estimated loss of $\le$1\% of the $S1$ signal due to late {\it phel}'s going under detection threshold. Switching to DC coupling was indeed motivated to favor a full photo-electron detection though at the expenses of a slightly higher dark current noise visible in the SER spectra and resulting in a lower peak-to-valley ratio.} 
and biased at negative voltage. After a period left for thermalization of the PMTs at LAr temperature (several hours),
the bias voltage on the PMTs cathode was slowly raised up to the working point HV= $-$1400 V corresponding to a gain of about 3$\times 10^6$.

\subsection{Data Analysis and Results}
\label{sec:4PMTtest}
The detector was mainly exposed to the $^{241}$Am source.
Exposures to $^{133}$Ba, $^{57}$Co and $^{137}$Cs sources (with $\gamma$ emission lines at higher energies) were also performed during the run period.  
The source was located inside a collimator holder positioned outside the vessel in a fixed position.
Data acquisition runs with the sources were alternated with blank runs (background from ambient radiation).

 The four signal waveforms were individually recorded for scintillation events in which the pulse from three of the four PMTs was above a threshold corresponding to 1.5 {\it phel}.  \\
During each source (or blank) run, single photo-electron pulses were selected in order to provide the SER data needed for calibration.\\
\begin{figure}[htbp]
\begin{center}
\vspace{-0.8cm}
\includegraphics[width=11.cm]{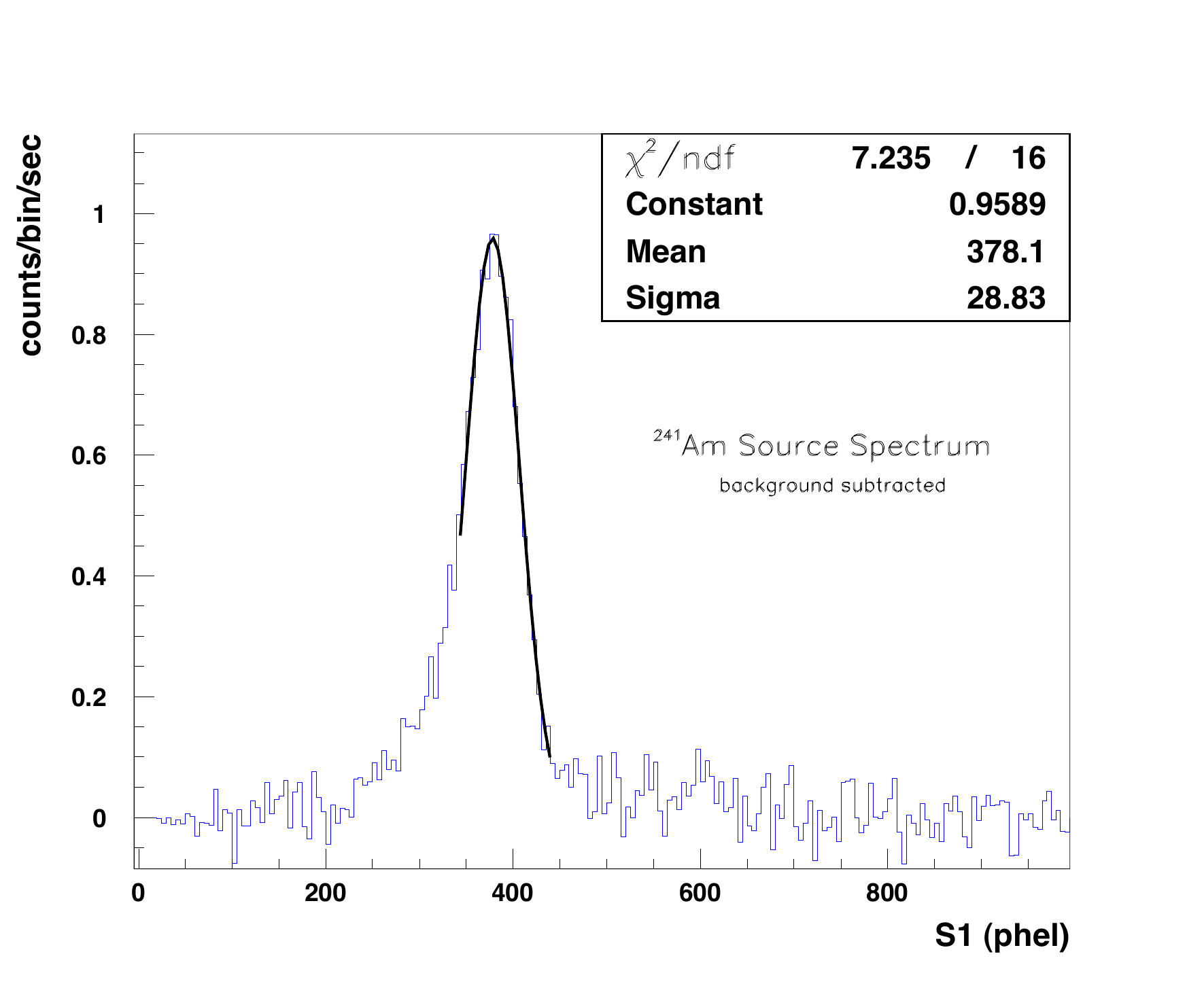}
\vspace{-0.6 cm}
\caption{\textsf{\textit{Four Hamamatsu PMTs test with the ``WArP-2.3 prototype": $^{241}$Am spectrum (background subtracted). Peak value and width determined by a gaussian fit (superimposed).
The relative energy resolution at the peak energy is $\frac{\sigma_E}{E} \simeq$ 8\%.}}}
\label{fig:Run12343_mod}
\end{center}
\end{figure}

It is worth noting that the detector geometry is a scaled-down version of the \WArP-100 detector (100 lt of active volume, 37 PMTs, $\sim$ 12\% of photo-cathodic coverage);  therefore the light yield from this detector test can be assumed as somehow predictive of the LY from the \WArP-100 Inner Detector, when operated under equivalent conditions.\\

A lot of attention has been given to the quality of the TPB coating on the VIKUITI ESR reflector. Before inserting it into the detector, the wavelength-shifting efficiency of the TPB coated reflector sheets was measured using a dedicated setup. Only after the expected maximum light output was confirmed the sheets were lined onto the internal surfaces of the chamber.\\


The analysis was performed in analogous way to the 2 PMT test, only this time the event signal amplitude was obtained by summing the single channel signal amplitudes after their normalization to the respective SER peak positions: ($S1=\sum_{i=1,..,4} s1_i$ in {\it phel} units;

An example of a Pulse Amplitude spectrum ($S1$ distribution) from the $^{241}$Am (higher intensity) source  
 is shown in Fig.\ref{fig:Run12343_mod} (background subtracted), with the full absorption peak at 378 {\it phel} as determined by a gaussian fit of the spectrum. 
Therefore, the Light Yield of the detector
can be evaluated as:\\
\begin{equation}
LY~=~6.35~\frac{phel}{\rm{keV}}~\pm~5\%
\label{eq:LY}
\end{equation}

The LY value determined with Compton spectra obtained from exposures to the other sources at higher $\gamma$-energies was less precise, but in good agreement with the above LY value.\\
\begin{figure}[h]
\begin{center}
\vspace{-0.6 cm}
\includegraphics[width=11.cm]{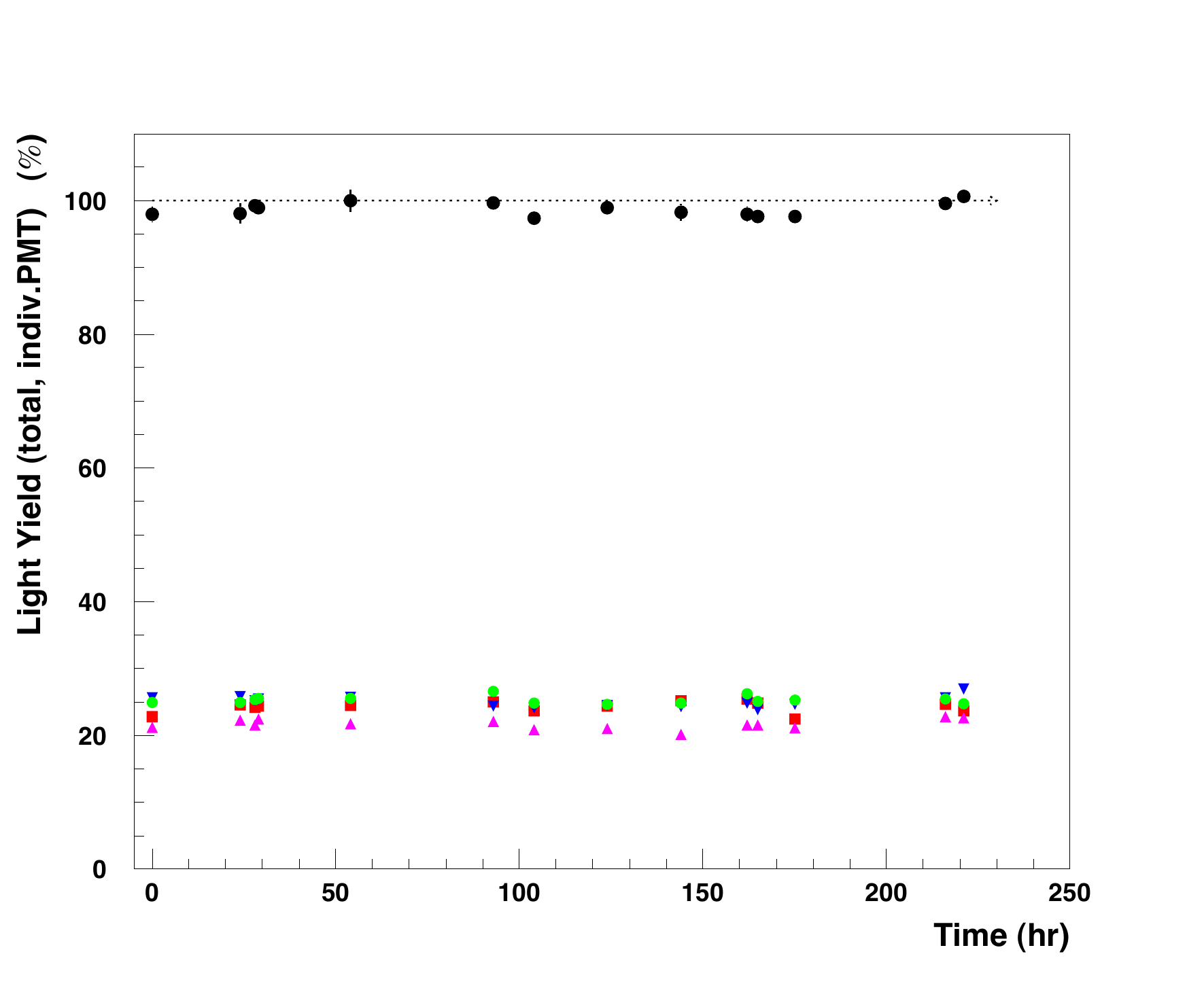}
 \vspace{-0.3 cm}
\caption{\textsf{\textit{Four Hamamatsu PMT test with the ``WArP-2.3 prototype": Light Yield variation over time and relative contributions from individual PMT.}}}
\label{fig:tot_rel-ly_vs_time}
\end{center}
\end{figure}
The LY stability in time was monitored over several time intervals during the test period. As an example, results from a ten-days stability test is shown in the Fig.\ref{fig:tot_rel-ly_vs_time}.  The LY (sum of all 4 PMTs) is stable within 2\%. \\
The LY from the individual PMTs was also checked by fitting the $s1_i$ distributions and the results (relative LY contribution) are shown in Fig.\ref{fig:tot_rel-ly_vs_time}. Three of the four PMTs behave in a very similar manner, while the fourth of these shows a systematically slightly lower value, not compatible with the difference in Quantum Efficiency.
 The reasons of this effect have yet to be understood. This is the only sign of malfunctioning experienced during the extensive use of the new Hamamatsu R11065 PMTs (it is worth noting that with this PMTs working at its nominal performance, i.e. like the other three PMTs, a LY $\simeq$ 6.6 phel/keV could be achieved).\\
\begin{figure}[h]
\begin{center}
\vspace{-0.6 cm}
\includegraphics[width=10.cm, height=9.cm]{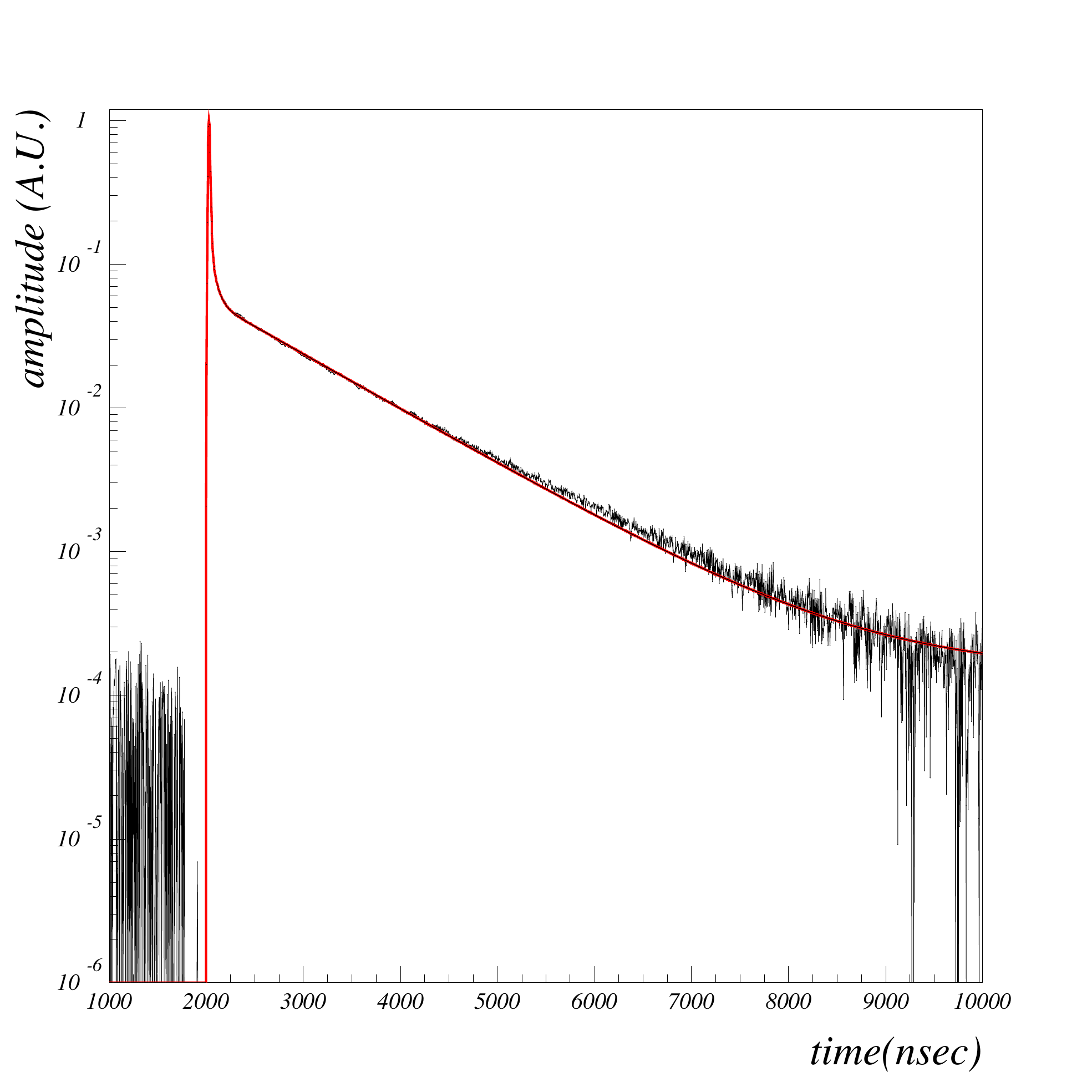}
 \vspace{-0.3 cm}
\caption{\textsf{\textit{Four Hamamatsu PMTs test with the ``WArP-2.3 prototype": Average waveform and exponential fit for $\tau_T$ determination.}}}
\label{fig:avg-wfm-FebMar11test}
\end{center}
\end{figure}

The purity of the liquid Argon during the test was inferred from the measurement of the long-decay time constant ($\tau_T$) of the scintillation light. The waveform from the sum of the four PMTs signals averaged over a large number of scintillation events is shown in Fig.\ref{fig:avg-wfm-FebMar11test}. The decay time of the slow component  from the fit was $\tau_T$  = 1130 ns.  This value - compared to an expected value around 1300 ns for asymptotically pure liquid Argon \cite{heindl,warp_N2} -  indicates that a residual concentration of impurities was still present in the liquid. \\
A direct measurement by mass spectroscopy on an Ar sample extracted from the chamber has been performed and showed the presence of Nitrogen in the liquid at the ppm level. Nitrogen impurities are not filtered out with the implemented set of filters (dedicated to O$_2$ and H$_2$O removal). Commercial Ar (research grade - routinely used for filling) is usually provided with a lower content of N$_2$ impurities, however a more dirty batch of Argon delivered in occasion of the present  test and used for the filling cannot be excluded. 
The reduction of the long-decay time constant  (via quenching effect of the Ar$_2^*$ excimers in triplet state) results in a $\sim$10 \% loss of light \cite{warp_N2} (and correspondingly in the LY value).

Therefore, the result with this  4.3 lt chamber (4 HQE Hamamatsu PMTs) agrees in good approximation with the result obtained with the 0.5 lt detector  equipped with one HQE PMT (LY~$\simeq$~7~{\it phel}/keV) and characterized by an equivalent photo-cathodic coverage ($\sim$12\% in both chambers). The difference in measured values can be attributed in full to a higher N$_2$ concentration in the second measurement. \\

\section{Conclusions}

A new PMT type with enhanced Quantum Efficiency photocathode and operating at LAr temperature has been recently developed by  {\sf Hamamatsu Photonics} Mod. R11065 with peak QE up to about 35\%. This PMT is very interesting from the point of view of Liquid Argon based Dark Matter detectors, which currently implement photo-multiplier tubes for signal read-out.  
 The new PMTs have been extensively tested in the course of the R\&D program of the \WArP Collaboration.\\
The main working parameters of this PMT were measured at LAr temperature and its optimal performance  has been demonstrated. \\
It has also been shown experimentally that 
Liquid Argon detectors with HQE photo-cathodic coverage in the 12\% range can achieve a light yield around 7 {\it phel}/keV (at null electric field), sufficient for detection of events down to few keV of energy deposition in the liquid.
However, since the scaling-up of detector mass and complexity (e.g. increased number of PMTs) without loss of detector performance (namely LY) is definitively not a trivial task, a dedicated test with four PMTs viewing signals from a 4.3 liters LAr cell has been performed. The main delicate variables to keep strictly under control in order to obtain a high Light Yield have been identified: (1) PMTs at the highest possible level of performance (Quantum Efficiency of the photo-cathode at LAr temperature, Collection Efficiency at the first dynode, overall PMT stability in response), (2) wavelength-shifting efficiency  of the TPB coating, (3) purity of the liquid Argon. \\
All three elements need to be simultaneously controlled and maintained at their best possible level to guarantee  optimal detector performance in line with expectations.


\end{document}